\documentclass[%
 aip,
rsi,%
 amsmath,amssymb,reprint
]{revtex4-1}

\usepackage{graphicx}
\usepackage{dcolumn}
\usepackage{bm}

\usepackage[breaklinks=true]{hyperref}

\begin{document}

\preprint{AIP/123-QED}

\title{Infrared nanoscopy down to liquid helium temperatures}

\author{Denny Lang}
\email{d.lang@hzdr.de}
\affiliation{Helmholtz-Zentrum Dresden-Rossendorf, Institute of Ion Beam Physics and Materials Research, 01328 Dresden, Germany.}
\affiliation{Institute of Applied Physics, Technische Universit\"at Dresden, 01062 Dresden, Germany.}

\author{Jonathan D\"oring}
\thanks{contributed equally to this work}
\affiliation{Institute of Applied Physics, Technische Universit\"at Dresden, 01062 Dresden, Germany.}

\author{Tobias N\"orenberg}
\affiliation{Institute of Applied Physics, Technische Universit\"at Dresden, 01062 Dresden, Germany.}

\author{\'{A}d\'{a}m Butykai}
\affiliation{Department of Physics, Budapest University of Technology and Economics and MTA-BME Lend\"ulet Magneto-Optical Spectroscopy Research Group, 1111 Budapest, Hungary.}

\author{Istv\'{a}n K\'{e}zsm\'{a}rki}
\affiliation{Department of Physics, Budapest University of Technology and Economics and MTA-BME Lend\"ulet Magneto-Optical Spectroscopy Research Group, 1111 Budapest, Hungary.}
\affiliation{Experimental Physics V, Center for Electronic Correlations and Magnetism, Institute of Physics, University of Augsburg, 86135 Augsburg, Germany.}

\author{Harald Schneider}
\affiliation{Institute of Ion Beam Physics and Materials Research, Helmholtz-Zentrum Dresden-Rossendorf, 01328 Dresden, Germany.}

\author{Stephan Winnerl}
\affiliation{Institute of Ion Beam Physics and Materials Research, Helmholtz-Zentrum Dresden-Rossendorf, 01328 Dresden, Germany.}

\author{Manfred Helm}
\affiliation{Institute of Ion Beam Physics and Materials Research, Helmholtz-Zentrum Dresden-Rossendorf, 01328 Dresden, Germany.}
\affiliation{Institute of Applied Physics, Technische Universit\"at Dresden, 01062 Dresden, Germany.}
\affiliation{cfaed $-$ Center for Advancing Electronics Dresden, Technische Universit\"at Dresden, 01062 Dresden, Germany.}

\author{Susanne C. Kehr}
\affiliation{Institute of Applied Physics, Technische Universit\"at Dresden, 01062 Dresden, Germany.}

\author{Lukas M. Eng}
\affiliation{Institute of Applied Physics, Technische Universit\"at Dresden, 01062 Dresden, Germany.}
\affiliation{cfaed $-$ Center for Advancing Electronics Dresden, Technische Universit\"at Dresden, 01062 Dresden, Germany.}

\date{\today}

\begin{abstract}
We introduce a scattering-type scanning near-field infrared microscope (s-SNIM) for the local scale near-field sample analysis and spectroscopy from room (RT) down to liquid helium (LHe) temperatures.  The extension of s-SNIM down to $T=5\,\text{K}$ is in particular crucial for low-temperature phase transitions, e.g. for the examination of superconductors, as well as low energy excitations. The LT s-SNIM performance is tested with CO$_2$-IR excitation at $T  =  7\, \text{K}$ using a bare Au reference and a structured $\text{Si/SiO}_2$-sample. Furthermore, we quantify the impact of local laser heating under the s-SNIM tip apex by monitoring the light-induced ferroelectric-to-paraelectric phase transition of the skyrmion-hosting multiferroic material GaV$_4$S$_8$ at $T_\text{c}=42\, \text{K}$. We apply LT s-SNIM to study the spectral response of GaV$_4$S$_8$ and its lateral domain structure in the ferroelectric phase by the mid-IR to THz free-electron laser-light source FELBE at the Helmholtz-Zentrum Dresden-Rossendorf, Germany. Notably, our s-SNIM is based on a non-contact atomic force microscope (AFM), and thus can be complemented in-situ by various other AFM techniques, such as topography profiling, piezo-response force microscopy (PFM) and/or Kelvin-probe force microscopy (KPFM). The combination of these methods support the comprehensive study of the mutual interplay in the topographic, electronic and optical properties of surfaces from room temperature down to $5\,\textrm{K}$.
\end{abstract}

\keywords{near-field microscopy, free-electron laser, mid-infrared, low temperature, phase transition, scanning force micropscopy}
\maketitle

\section{\label{sec:level1}Introduction}

\begin{figure}
\centering
\includegraphics[width=0.7\linewidth]{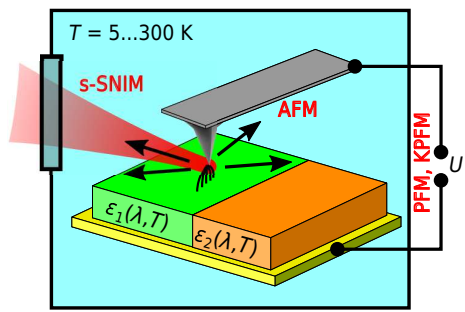}
\caption{\label{fig:teaser} Sketch of the LT s-SNIM: IR/THz radiation is focused onto the tip-sample junction inside the LHe cryostat. Elastically backscattered radiation from the tip-sample system is measured and the near-field contributions extracted, which contain the local information of the sample permittivity $\varepsilon$ as a function of temperature $T$ and wavelength $\lambda$. Both sample and tip are electrically contacted in order to enable complementary AFM-techniques such as PFM or KPFM.}
\end{figure}

In the last decade, scattering-type scanning near-field infrared optical microscopy (s-SNIM) has developed into a powerful and widespread method for the nanoscale imaging and spectroscopic analysis at infrared (IR) and terahertz (THz) wavelengths \cite{Hillenbrand2002,Huth2012,Winnerl2016,Huber2016,Amenabar2017}. Providing a resolution ways beyond the classical diffraction limit \cite{Knoll2000}, s-SNIM has valuably contributed to the fields of 2D materials (e.g. graphene \cite{Ni2016,Jiang2016,Alonso-Gonzalez2016} and MoS$_2$ \cite{Patoka2016}), to the research of metamaterials and superlenses \cite{Taubner2006,Kehr2011,Fehrenbacher2015}, as well as to the investigation of semiconductor and organic nanostructures \cite{Huth2012, Westermeier2014}, just to mention a few examples. Nevertheless, a lot of prominent effects in solid-state physics such as superconductivity \cite{Yamamoto2007, Ahmad2015}, phase transitions in multiferroic materials \cite{Doring2014,McLeod2016} or the confinement in low-dimensional quantum structures \cite{Pajot2010} need to be explored at low temperatures (LT). Combining LT with the huge potential of IR and THz nanoscopy thus promises great advances in understanding and probing fundamental excitations, such as phonons, magnons and polaritons, in these exotic material classes. The goal of this work is to advertise such a promising combination that involves the “FELBE” tunable free-electron laser (FEL) at the Helmholtz-Zentrum Dresden-Rossendorf, Germany, as a high repetition rate, narrow-band light source tunable from 5 to $250\,\mu$m wavelength.

To the best of our knowledge, the first work on LT s-SNIM was reported by Moldovan-Doye et al. \cite{Moldovan-Doyen2011} and used the s-SNIM to collect (rather than excite) mid-IR to THz photons with sub-diffraction resolution at $T = 100\,\text{K}$, exploring the light emission of an (active) quantum cascade laser element that was coupled to a photonic crystal. Moreover, Yang et al. \cite{Yang2013} demonstrated LT s-SNIM at $T = 200\, \text{K}$ on non-conducting domains in $\text{V}_2\text{O}_3$ single crystals that nucleate around topographic defects. Furthermore, a recent publication by McLeod et al. \cite{McLeod2016} impressively showed the detailed mapping of the $\text{V}_2\text{O}_3$ insulator-to-metal transition at $T > 160\, \text{K}$. 

Our work here demonstrates LT s-SNIM (Fig.~\ref{fig:teaser}) operation down to $5\, \text{K}$, while exploring different samples including Au, structured Si-SiO$_2$ and multiferroic $\textrm{GaV}_4\textrm{S}_8$ with both the CO$_2$ laser and the FEL \cite{Winnerl2016}. The overall LT s-SNIM functionality at various temperatures is illustrated by inspecting a topographically planar Au surface as well as a structured $\textrm{Si/SiO}_2$-sample. We study the multiferroic material $\textrm{GaV}_4\textrm{S}_8$ \cite{Milde2013,Wang2015,Kezsmarki2015,Butykai2017} in order to quantify the local impact of IR radiation by the tip-sample confinement. Moreover the local spectral response as well as the lateral distribution of structural domains below $42\,\text{K}$ are studied on the same $\textrm{GaV}_4\textrm{S}_8$ sample by means of in-situ LT s-SNIM, PFM, and KPFM \cite{Doring2014,Doring2016a,Butykai2017}.

\section{\label{sec:level2}Setup}

\subsection{\label{sec:level21}s-SNIM}

In s-SNIM, infrared radiation is scattered by the AFM tip that oscillates a few nm above the sample surface. Due to the tip-sample interaction in the optical near-field, the radiation scattered to the far field contains information on the local refractive index tensor of the sample \cite{Knoll2000,Schneider2005}.

Our LT s-SNIM is based on a custom-made LT atomic force microscope (AFM) system (by \textit{Attocube Systems AG}) using a \textit{Janis Research Company} LHe bath cryostat. For s-SNIM measurements, the AFM is operated in tapping-mode \cite{Zhong1993}, with the cantilever being mechanically excited at its resonance frequency $f_0$ of about $170\,\text{kHz}$ with a typical oscillation amplitude of 50~nm. The actual cantilever motion is detected with a fiber-based homodyne interferometer at $\lambda=1310\,\text{nm}$. We use standard non-contact AFM cantilevers (\textit{PPP-NCLPt} from \textit{NANOSENSORS}) for all measurements. AFM tips possess a $25\,\text{nm}$ PtIr5 coating that enables spectrally flat optical excitation over the whole wavelength range from 5 to $250\,\mu \text{m}$ \cite{Rakic1998}. Equally, such tips show an excellent thermal and electrical conductivity, e.g.\ to allow for specific complementary AFM-mode operation such as PFM or KPFM at one and the same sample position.

PFM makes use of the converse piezoelectric effect, and thus is able to probe different components of the local piezoelectric tensor of the material under inspection \cite{Guthner1992,Eng1999,Doring2016a}. Applying a 1 - 10 kHz AC voltage with a typical amplitude of $1-10\,\text{V} $ between tip and sample in contact-mode of the AFM leads to local distortions and sample oscillations that are sensed by the AFM probe and demodulated via a lock-in amplifier. The lock-in amplifier measures both amplitude $A_\text{PFM}$ and phase $\Phi_\text{PFM}$ with respect to the excitation. Note that due to the interferometric cantilever detection, only out-of-plane components of the piezoelectric tensor can be addressed in our LT set-up.

KPFM, on the other hand, operates in non-contact mode, and allows to locally quantify and compensate any electrostatic potential offset between tip and sample surface \cite{Nonnenmacher1991,Zerweck2005}. KPFM hence not only minimizes cross-talk of electrostatic forces and topography or s-SNIM signals, but equally allows for probing the local work function \cite{Shusterman2007} or sample surface photo-voltage \cite{Schumacher2016} of the sample system simultaneously to s-SNIM investigations. For KPFM, we typically apply a 1 V AC-bias at 4 kHz to the tip on top of the fundamental oscillation amplitude of the cantilever.

\subsection{\label{sec:level22}LT s-SNIM optical setup}

\begin{figure}
\includegraphics[width=0.5\textwidth]{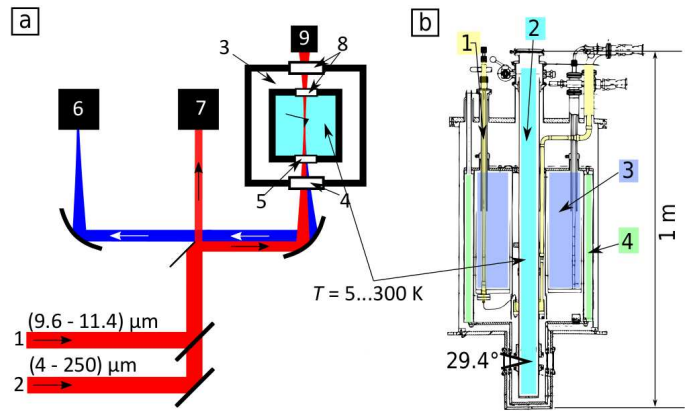}
\caption{\label{fig:setup} (a) Scheme of the optical setup, combined with the LT s-SNIM: the tunable infrared laser source [CO$_2$ laser (1) or FEL (2)] is focused on the tip-sample system in the cryostat (3) through an outer window (4; KRS5, quartz or ZnSe) and an inner diamond window (5). The backscattered radiation is again focused on and detected by a infrared/THz detector (6; MCT, Ge:Ga detector or hot-electron bolometer). The laser power is recorded simultaneously by a powermeter (7). The quartz windows (8) enable opical access for alignment and positioning by a camera (9). (b) Cryostat with heat-exchange system (1), sample tube (2) containing the AFM (not shown here), and LHe (3) and LN$_2$ (4) dewar vessels.}
\end{figure}

The optical setup of our s-SNIM is sketched in Fig.~\ref{fig:setup} (a). As shown, either a narrow-band table-top $\text{CO}_2$ laser or the FEL at Helmholtz-Zentrum Dresden-Rossendorf are used as IR excitations. The FEL wavelength is tunable over a broad range from $5\,\mu\text{m}$ to $250\,\mu\text{m}$ ($2000\,\text{cm}^{-1}$ to $40\,\text{cm}^{-1}$,  $60\,\text{THz}$ to $1.2\,\text{THz}$ or $248\,\text{meV}$ to $5\,\text{meV}$), and provides a 1-25 ps pulse train at a repetition rate of 13 MHz. The typical spectral width is about $1\, \%$ of the actual wavelength. The tunability of the continuous wave (cw) $\text{CO}_2$ laser is limited to various sharp lines between $9.6\, \mu\text{m}$ to $11.4\, \mu\text{m}$.

After passing a geometric beam-splitter (BS), half of the incident radiation is directed to a power meter to monitor the stability of the incident beam at a typical power of $10$ to $50 \, \text{mW}$ at the position of the LT s-SNIM probe. The other part of the beam is focused onto the tip apex using a $3^{\prime\prime}$ Au-coated parabolic mirror sitting outside of the cryostat, with a resulting numerical aperture of about 0.2 and a focus diameter of about $100\, \mu\text{m}$ at the s-SNIM probe for $\lambda=10\,\mu\text{m}$. Both the incident and the backscattered radiation pass through two cryostat windows: an outer window made of KRS-5, ZnSe or Quartz for different wavelength regimes, respectively, and an inner diamond window. This results in a total attenuation of about $50\,\%$ for every direction.  For all measurements presented here, the radiation is polarized parallel to the incident plane (p-pol) of the tip-sample system, which results in the strongest near-field response \cite{Knoll2000}. Nevertheless, it is possible to rotate the polarization by $90^\circ$ resulting in perpendicular (s-pol) polarization with respect to the incident plane.

The backscattered radiation, that carries the information about the refractive index of the sample on the local scale, is collected by the same parabolic mirror and, after passing through the BS, is focused onto the detector. Depending on the wavelength, we use either a photoconductive mercury cadmium telluride (MCT) detector ($4 \, \mu\text{m} <\lambda <  26 \, \mu \text{m}$), a LHe-cooled gallium doped germanium (Ge:Ga) photoconductive detector ($24 \, \mu\text{m} <  \lambda < 50 \, \mu \text{m}$), or a LHe-cooled indium anitmonide (InSb) hot-electron bolometer ($50\, \mu\text{m} <  \lambda < 250 \, \mu \text{m}$). The detector output is demodulated at the $n$-th harmonic of the cantilever resonance frequency $\Omega$ using a lock-in amplifier. The resulting signals are further referred to as $\textit{NF}_{n \Omega}$. This technique effectively suppresses far-field components in the detected signal \cite{Knoll2000,Wurtz1998,Hillenbrand2000}.

\subsection{\label{sec:level23}Cryostat}

The bath cryostat [see Fig.~\ref{fig:setup} (b)] consists of two dewar vessels for liquid nitrogen and liquid helium allowing for low-temperature measurements down to $T=5\,\text{K}$. The inner tank connects to the sample tube via a heat exchanger, where the liquid helium cooles the outer wall of the sample tube. The sample tube hosts the AFM and is filled with He gas at a pressure of about $100 \, \text{mbar}$ in order to facilitate heat transfer between the sample and the wall. 

Two thermosensors combined with heating elements are located directly below the sample holder as well as on the wall of the sample tube at the height of the tip. Whenever the temperature at both thermosensors are equal, the system is in a thermal equilibrium, which is the starting point for all LT s-SNIM investigations. The temperature measured at the sample holder is hereafter labeled as $T_s$, whereas the actual temperature on the sample beneath the tip can differ due to local laser heating as discussed later. The temperature $T_s$ is controlled by the flow of liquid gas and a temperature controller connected to both heating elements.

The optical access to the cryostat is provided in two separate paths: the near-field s-SNIM path (excitation and detection) is coupled in via the parabolic mirror and is focused onto the AFM tip as described above, while the monitoring and adjustment path hosts an optical camera mounted at the rear exit of the cryostat (see Fig.~\ref{fig:setup}). 

\section{\label{level3}Results}

\subsection{\label{sec:level31}LT s-SNIM perfomance}

\begin{figure*}
\includegraphics[width=0.75\textwidth]{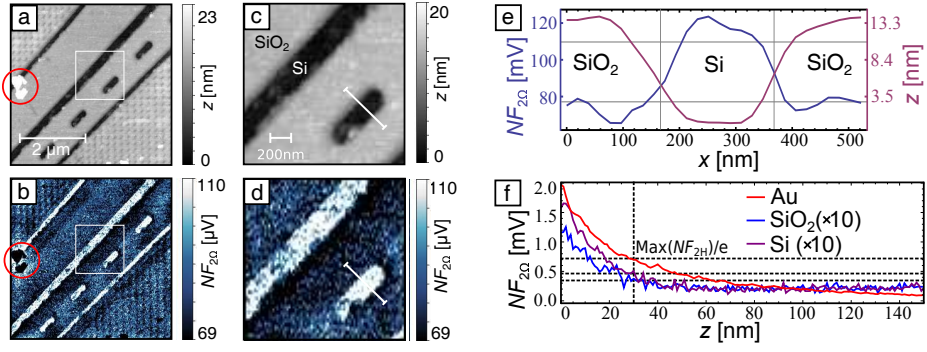}
\caption{\label{fig:Silicon}Topography (a) and near-field (b) scans at $T_\text{s}=7\,\text{K}$ of structured Si-$\text{SiO}_2$ with associated zoom areas (c) and (d). The corresponding profiles (e) demonstrate a spatial LT s-SNIM resolution better than $100\,\text{nm}$. The measured retract curves (f) of Au (reference), Si and $\text{SiO}_2$ confirm the near-field characteristics \cite{Knoll2000} of the measurements with a non-linear decay of $\textit{NF}_{2 \Omega}$ within about $30 \, \text{nm}$ at $\text{Max}(\textit{NF}_\text{2H})/e$.}
\end{figure*}

We verify the near-field perfomance of our LT s-SNIM system by probing a structured $\text{Si/SiO}_2$-sample (for details see Apendix A) at a nominal temperature of $T_\text{s}=7 \, \text{K}$ using the $\text{CO}_2$ laser at a wavelength of $\lambda = 9.7 \, \mu \text{m}$ (see Fig.~\ref{fig:Silicon}). Note that this sample system has been chosen deliberately to reveal a s-SNIM contrast that is temperature independent and thus can be compared with RT results \cite{Kehr2017}.

A topography scan of this sample is depicted in Fig.~\ref{fig:Silicon} (a) with a zoomed-in sector (white rectangle) shown in Fig.~\ref{fig:Silicon} (c). The higher, bright areas represent $\text{SiO}_2$ whereas the lower, dark areas correspond to Si. The scanned section contains stripes and rectangles of different sizes. Note that the smallest of them are even on the magnitude of the probe's apex diameter of about $50\,\text{nm}$. We obtain a RMS-roughness of $0.5 \, \text{nm}$ on $\text{SiO}_2$, which proves the excellent AFM stability at low temperatures. The corresponding, simultaneously recorded near-field signal [Fig.~\ref{fig:Silicon} (b) with zoom (d)] reveals an inverted optical contrast with respect to the topography: Here the  $\text{SiO}_2$ areas appear darker than the Si areas. This can be explained by the values of the wavelength-dependent permittivities of both materials taking $\varepsilon_\text{Si}=11.8+3\cdot10^{-4}\text{i}$ \cite{Chandler-Horowitz2005} and  $\varepsilon_{\text{SiO}_2}=6.4+7.9\text{i}$ \cite{Kischkat2012} for $\lambda=9.7 \, \mu \text{m}$ resulting in lower and higher s-SNIM signals, respectively \cite{Knoll2000}. The signal-to-noise ratio of $\textit{NF}_{2 \Omega}$ for $\text{SiO}_2$ is approximately 9 and even the smallest structures are clearly visible (top-left and bottom-right areas). Residual particles on the scan, that can neither be assigned to $\text{SiO}_2$  nor to Si, appear dark in the near-field image [see red circle in Fig.~\ref{fig:Silicon} (a) and (b)]. This indicates that they are most likely highly insulating and being nonresonant with respect to the excitation wavelength.

A more detailed view is presented in the cross section profile in Fig.~\ref{fig:Silicon} (e). The  $\text{Si}$-$\text{SiO}_2$ edge in the topography shows a width of less than $100\,\text{nm}$ in the AFM. Notably, the spatial resolution of $\textit{NF}_{2 \Omega}$ is similar. To verify the near-field character of  $\textit{NF}_{2 \Omega}$, we compare its dependence on the tip-sample distance by retract curves on Si, $\text{SiO}_2$, and Au, respectively, as depicted in Fig.~\ref{fig:Silicon} (f). Both for Si and $\text{SiO}_2$, \textit{NF}$_{2 \Omega}$ is more than 10 times smaller than for the plane Au reference sample since Au, as a metal is highly reflective ($\varepsilon_\text{Au}=-4330+757\text{i}$ \cite{Babar2015}), which results in a huge near-field signal. On all three materials, a non-linear decay of $NF_{2 \Omega}$, typical for the near field, is observed within $30 \, \text{nm}$ at $\text{Max}(\textit{NF}_\text{2H})/e$. This decay is known to scale with the tip diameter \cite{Knoll2000} and therefore is in good agreement with the observed spatial resolution better than $100\,\text{nm}$.

\subsection{\label{sec:level32}Temperature calibration}

\begin{figure}
\includegraphics{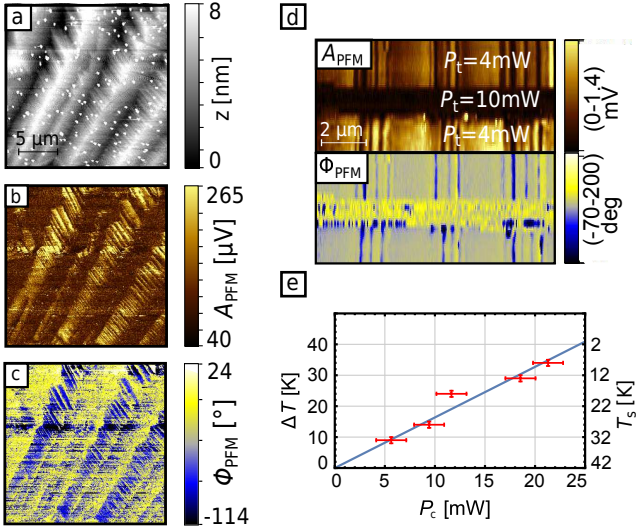}
\caption{\label{fig:GVS} Scan on the (111) surface of a GaV$_4$S$_8$ single crystal: (a) topography, (b) PFM amplitude $A_\text{PFM}$, and (c) PFM phase $\Phi_\text{PFM}$, displaying the lamella-shaped ferroelectric domain structure at $T_\text{s}=20\,\text{K}$. (d) An exemplarily PFM line scan over domains two different laser powers reveals the dis- and reappearing of ferroelectric domains at $T_\text{s}=35\,\text{K}$. (e) Linear correlation between heating $\Delta T$ and critical laser power $P_\text{c}$.   }
\end{figure}

The sample temperature $T_\text{s}$ of our LT s-SNIM system is measured below the sample holder and is thus accurate for samples with excellent thermal conductivity like metals. However, the temperature of the sample at the local position of the tip may be higher due to local heating by the focused IR-laser and additionally due to the field enhancement at the tip apex of the metallic probe. This heating effect is expectedly influenced by sample parameters such as thermal conductivity and wavelength-dependent infrared absorption. Especially the latter will play an important role for the s-SNIM investigation of materials close to absorption resonances, for instance within the Reststrahlenband. Measurements in this regime are of particular interest for s-SNIM since the near-field enhancement conditions result in an enhanced sample sensitivity \cite{Hillenbrand2002,Kehr2008}. 

In order to determine experimentally the local sample heating at the tip induced by the infrared laser, the Jahn-Teller transition in multiferroic $\text{GaV}_4\text{S}_8$ is used \cite{Wang2015,Butykai2017}. This material becomes ferroelectric below a critical temperature $T_\text{c} \, =\, 42\,\text{K}$ \cite{Butykai2017}, which induces a clear contrast in the PFM-signal (see Fig.~\ref{fig:GVS} d) due to ferroelectric domains with polarizations pointing along any of the four $\langle111\rangle$-type axes.  Fig.~\ref{fig:GVS} (a)-(c) shows topography and PFM images of the (111) surface of a $\text{GaV}_4\text{S}_8$ single crystal at $T_s\,=20\,\text{K}$, which are all clearly display the lamellar ferroelectric domain pattern. We measure the out-of-plane piezoelectric tensor components discussed in detail by Butykai et al. \cite{Butykai2017}.

We study the effect of local heating with a $\text{CO}_2$-laser at $10.6\,\mu \text{m}$. Starting with various temperatures $T_\text{s} \, <\, T_\text{c}$ the sample is locally heated by the $\text{CO}_2$-laser. We monitor the temperature $T_\text{s}$ by the two thermosensors of the cryostat starting in thermal equilibrium. Here, a 2D image of PFM amplitude ($A_\text{PFM}$) and PFM phase ($\Phi_\text{PFM}$) of the $\text{GaV}_4\text{S}_8$ sample is recorded to obtain the domain pattern and to chose a suitable line with clear domain contrast [Fig.~\ref{fig:GVS} (d)]. During calibration, this line is scanned repeatedly while the laser power $P_\text{t}$  is increased gradually until the domains disappear in the PFM-signals at a critical laser power $P_\text{c}$. For $P_\text{t}>P_\text{c}$ the sample is locally heated up to the paraelectric phase, hence the domains vanish leading to zero $A_\text{PFM}$ and random $\Phi_\text{PFM}$ signals. At the transition point, the local temperature due to laser heating reaches $T_\text{c}$. Consequently, the difference $\Delta T \,=\,T_\text{c}\,-\, T_\text{s}$ is the resulting temperature increase due to power-dependent sample heating. Fig.~\ref{fig:GVS} (d) exemplarily displays such line scans with vanishing domains upon IR illumination. We note that the reappearing domain pattern is probably highly influenced by the ferroelectric boundaries of the heated paraelectric spot. This laser-induced local heating therefore often fully reproduces or strongly correlates with the initial domain structure, depending on the size of the spot heated above the transition temperature. The systematic analysis of controlled heating experiments will be the aim of a future publication. By setting different starting temperatures $T_\text{s}$ we find a linear dependence of $\Delta T$ on $P_\text{c}$ for the measured temperature range and extract a laser heating coefficient $\xi \,=\,\Delta T/P_\text{c}\,=\,(1.64\pm0.1) \,\frac{\text{K}}{\text{mW}} $ [see Fig.~\ref{fig:GVS} (e)]. Even though this coefficient depends on material parameters in terms of heat conductivity and absorbance it gives a general estimation of heating effects in our specific LT s-SNIM setup for off-resonant heating far from absorption peaks.

In order to validate the heating coefficient for resonant excitation close to the Reststrahlenband we have applied the same method while operating the FEL around $31.7 \, \mu \text{m}$. In this case, $\xi$ is about 10 to 15 \% higher as compared to the off-resonant case even though the focus diameter is about 3 times larger than with the CO$_2$ laser at $10.6\,\mu\text{m}$. However, no significant wavelength-dependence could be observed for the resonant case, indicating a major impact of the specific s-SNIM geometry including the metal-coated tip and cantilever on the heating, rather than absorption by the sample.

\begin{figure*}
\includegraphics{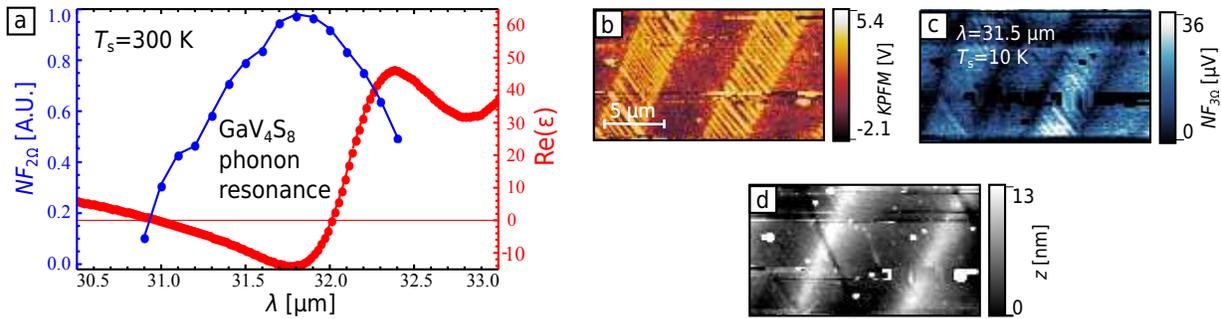}
\caption{\label{fig:GVS3}Near-field study of GaV$_4$S$_8$: (a) The near-field signal $\textit{NF}_{2\Omega}$ (blue) shows a distinct phonon resonance within the Reststrahlenband at about $31.8\,\mu\text{m}$ at $T_\text{s}=300\,\text{K}$. The position of the resonance matches well to the permittivity $\varepsilon$ measured by Reschke et al. \cite{Reschke2017} with a negative $\text{Re}(\varepsilon)$. In the LT-phase at $T_\text{s}=10\,\text{K}$, below the critical temperature, the lamella-shaped domain structure appears both in the KPFM (b) and s-SNIM (c) signals by scanning the (111) surface topography (d) of  GaV$_4$S$_8$.}
\end{figure*}

\subsection{\label{sec:level33}Near-field examination of GaV$_4$S$_8$ at $T_\text{s}=10\,\text{K}$}

We study the local optical response of the GaV$_4$S$_8$ (111) surface in the ferroelectric phase below the Jahn-Teller transition at $T_\text{c}=42\,\text{K}$. Note that GaV$_4$S$_8$ is optically anisotropic where the optical axes of the different domains are parallel to the $\langle111\rangle$-type directions. It has been shown earlier that such anisotropic domains may be probed in resonant s-SNIM  \cite{Doring2014,Kehr2008} at specific wavelengths within the Restrahlenband of the material, making use of the strong IR-active phonon $\text{F}_2$ at a wavelength of about $32\,\mu\text{m}$ or $310\,\text{cm}^{-1}$ \cite{Hlinka2016,Reschke2017}. 

For  GaV$_4$S$_8$, such a resonance in the near field occurs at around $\lambda=31.8\,\mu\text{m}$ resulting in an enhanced s-SNIM signal as shown in Fig.  \ref{fig:GVS3} (a) for a temperature of $T_\text{s}=300\,\text{K}$. At the same wavelength the permittivity $\text{Re}(\varepsilon)$ gets negative \cite{Reschke2017} which is crucial for observing a strong resonance in the near-field signal \cite{Kehr2008}. Probing the sample surface within the LT- phase, we observe the typical lamella-shaped domain structure \cite{Butykai2017} in all in-situ accessible channels for non-contact AFM, i.e.\ topography, KPFM, and near-field response [Figs. \ref{fig:GVS3} (b)-(d)]. Considering the laser power of $P_\text{t}=12\,\text{mW}$, the temperature calibration discussed above determines the local sample temperature of around $30\,\text{K}$ within the ferroelectric phase below $T_\text{c}=42\,\text{K}$, which is confirmed by all measured channels, topography, s-SNIM and KPFM.

Please note that we have deliberately chosen here a structure that is known to give the same results for all channels. However, as completely different sample properties are probed with AFM, s-SNIM, and KPFM, namely topography, local infrared-optical response and surface potential, these techniques are complementary and in other types of samples a full comprehensive study is enabled.

\section{\label{level4}Conclusion}

We have reported here a LT s-SNIM system that, in combination with the FEL as a spectrally narrow, tunable IR laser light source, is capable of operating down to $5 \, \text{K}$ over the broad wavelength range of 5 to 250 $\mu \text{m}$. The LT s-SNIM allows for sub-diffraction limited infrared optical examination with a wavelength-independent lateral resolution of about $50\, \text{nm}$ ($\lambda/2000$ at $10\,\mu\text{m}$). We demonstrate LT s-SNIM operation with respect to distance decay on a Au-film, lateral resolution on Si-SiO$_2$ and spectral as well as structural resolution on GaV$_4$S$_8$. Furthermore, we have calibrated the temperature impact due to local laser heating using the Jahn-Teller phase transition of GaV$_4$S$_8$. Moreover, the setup supports complementary AFM techniques such as KPFM and PFM. The combination of these measurement techniques enables far-reaching opportunities for detailed investigations of phase transitions, superconducting materials and quantum structures on the nm length scale.  
\begin{acknowledgments}

The authors are greatful to the ELBE team at the Helmholtz-Zentrum Dresden-Rossendorf for the operation of the free-electon laser FELBE and for dedicated support. We thank Stephan Reschke as well as Franz Mayr from the Insitute of Physics, University of Augsburg, Germany for the permittivity data of GaV$_4$S$_8$. The study of GaV$_4$S$_8$ has been conducted within the framework of the Collaborative Research Center 'Correlated Magnetism: From Frustration to Topology’ (SFB 1143) via TP C05. We acknowledge the funding via BMBF Grants Nos. '05KI0ODB', '05KI0BRA' and '05K16ODA' as well as DFG No. 'KE2068/2-1'. The work was supported by the Cluster of Excellence 'Center for Advancing Electronics Dresden (cfaed)'. The work was supported by the joined DAAD project No. 'TKA-DAAD 152294' and also by the Deutsche Forschungsgemeinschaft(DFG) via the Transregional Research Collaboration TRR 80: From Electronic Correlations to Functionality (Augsburg - Munich - Stuttgart).

\end{acknowledgments}

\appendix

\section{$\text{Si/SiO}_2$-sample}

The structure was first etched into Si covered by a lithographic mask. Subsequently, the trenches were filled with amorphous SiO$_2$ by CVD-growth. After removing the lithography mask, the sample was polished resulting in a structure with only small topography variations between Si and SiO$_2$ and a thickness of the embedded SiO$_2$ of about $400\,\text{nm}$.

\bibliography{LT_v0}

\begin{thebibliography}{44}%
\makeatletter
\providecommand \@ifxundefined [1]{%
 \@ifx{#1\undefined}
}%
\providecommand \@ifnum [1]{%
 \ifnum #1\expandafter \@firstoftwo
 \else \expandafter \@secondoftwo
 \fi
}%
\providecommand \@ifx [1]{%
 \ifx #1\expandafter \@firstoftwo
 \else \expandafter \@secondoftwo
 \fi
}%
\providecommand \natexlab [1]{#1}%
\providecommand \enquote  [1]{``#1''}%
\providecommand \bibnamefont  [1]{#1}%
\providecommand \bibfnamefont [1]{#1}%
\providecommand \citenamefont [1]{#1}%
\providecommand \href@noop [0]{\@secondoftwo}%
\providecommand \href [0]{\begingroup \@sanitize@url \@href}%
\providecommand \@href[1]{\@@startlink{#1}\@@href}%
\providecommand \@@href[1]{\endgroup#1\@@endlink}%
\providecommand \@sanitize@url [0]{\catcode `\\12\catcode `\$12\catcode
  `\&12\catcode `\#12\catcode `\^12\catcode `\_12\catcode `\%12\relax}%
\providecommand \@@startlink[1]{}%
\providecommand \@@endlink[0]{}%
\providecommand \url  [0]{\begingroup\@sanitize@url \@url }%
\providecommand \@url [1]{\endgroup\@href {#1}{\urlprefix }}%
\providecommand \urlprefix  [0]{URL }%
\providecommand \Eprint [0]{\href }%
\providecommand \doibase [0]{http://dx.doi.org/}%
\providecommand \selectlanguage [0]{\@gobble}%
\providecommand \bibinfo  [0]{\@secondoftwo}%
\providecommand \bibfield  [0]{\@secondoftwo}%
\providecommand \translation [1]{[#1]}%
\providecommand \BibitemOpen [0]{}%
\providecommand \bibitemStop [0]{}%
\providecommand \bibitemNoStop [0]{.\EOS\space}%
\providecommand \EOS [0]{\spacefactor3000\relax}%
\providecommand \BibitemShut  [1]{\csname bibitem#1\endcsname}%
\let\auto@bib@innerbib\@empty
\bibitem [{\citenamefont {Hillenbrand}, \citenamefont {Taubner},\ and\
  \citenamefont {Keilmann}(2002)}]{Hillenbrand2002}%
  \BibitemOpen
  \bibfield  {author} {\bibinfo {author} {\bibfnamefont {R.}~\bibnamefont
  {Hillenbrand}}, \bibinfo {author} {\bibfnamefont {T.}~\bibnamefont
  {Taubner}}, \ and\ \bibinfo {author} {\bibfnamefont {F.}~\bibnamefont
  {Keilmann}},\ }\href {\doibase 10.1038/nature00899} {\bibfield  {journal}
  {\bibinfo  {journal} {Nature}\ }\textbf {\bibinfo {volume} {418}},\ \bibinfo
  {pages} {159} (\bibinfo {year} {2002})}\BibitemShut {NoStop}%
\bibitem [{\citenamefont {Huth}\ \emph {et~al.}(2012)\citenamefont {Huth},
  \citenamefont {Govyadinov}, \citenamefont {Amarie}, \citenamefont {Nuansing},
  \citenamefont {Keilmann},\ and\ \citenamefont {Hillenbrand}}]{Huth2012}%
  \BibitemOpen
  \bibfield  {author} {\bibinfo {author} {\bibfnamefont {F.}~\bibnamefont
  {Huth}}, \bibinfo {author} {\bibfnamefont {A.}~\bibnamefont {Govyadinov}},
  \bibinfo {author} {\bibfnamefont {S.}~\bibnamefont {Amarie}}, \bibinfo
  {author} {\bibfnamefont {W.}~\bibnamefont {Nuansing}}, \bibinfo {author}
  {\bibfnamefont {F.}~\bibnamefont {Keilmann}}, \ and\ \bibinfo {author}
  {\bibfnamefont {R.}~\bibnamefont {Hillenbrand}},\ }\href {\doibase
  10.1021/nl301159v} {\bibfield  {journal} {\bibinfo  {journal} {Nano Lett.}\
  }\textbf {\bibinfo {volume} {12}},\ \bibinfo {pages} {3973} (\bibinfo {year}
  {2012})}\BibitemShut {NoStop}%
\bibitem [{\citenamefont {Kuschewski}\ \emph {et~al.}(2016)\citenamefont
  {Kuschewski}, \citenamefont {von Ribbeck}, \citenamefont {D{\"{o}}ring},
  \citenamefont {Winnerl}, \citenamefont {Eng},\ and\ \citenamefont {{S. C.
  Kehr}}}]{Winnerl2016}%
  \BibitemOpen
  \bibfield  {author} {\bibinfo {author} {\bibfnamefont {F.}~\bibnamefont
  {Kuschewski}}, \bibinfo {author} {\bibfnamefont {H.-G.}\ \bibnamefont {von
  Ribbeck}}, \bibinfo {author} {\bibfnamefont {J.}~\bibnamefont
  {D{\"{o}}ring}}, \bibinfo {author} {\bibfnamefont {S.}~\bibnamefont
  {Winnerl}}, \bibinfo {author} {\bibfnamefont {L.~M.}\ \bibnamefont {Eng}}, \
  and\ \bibinfo {author} {\bibnamefont {{S. C. Kehr}}},\ }\href {\doibase
  10.1063/1.4943793} {\bibfield  {journal} {\bibinfo  {journal} {Appl. Phys.
  Lett.}\ }\textbf {\bibinfo {volume} {108}},\ \bibinfo {pages} {113102}
  (\bibinfo {year} {2016})}\BibitemShut {NoStop}%
\bibitem [{\citenamefont {Huber}\ \emph {et~al.}(2016)\citenamefont {Huber},
  \citenamefont {Mooshammer}, \citenamefont {Plankl}, \citenamefont {Viti},
  \citenamefont {Sandner}, \citenamefont {Kastner}, \citenamefont {Frank},
  \citenamefont {Fabian}, \citenamefont {Vitiello}, \citenamefont {Cocker},\
  and\ \citenamefont {Huber}}]{Huber2016}%
  \BibitemOpen
  \bibfield  {author} {\bibinfo {author} {\bibfnamefont {M.~A.}\ \bibnamefont
  {Huber}}, \bibinfo {author} {\bibfnamefont {F.}~\bibnamefont {Mooshammer}},
  \bibinfo {author} {\bibfnamefont {M.}~\bibnamefont {Plankl}}, \bibinfo
  {author} {\bibfnamefont {L.}~\bibnamefont {Viti}}, \bibinfo {author}
  {\bibfnamefont {F.}~\bibnamefont {Sandner}}, \bibinfo {author} {\bibfnamefont
  {L.~Z.}\ \bibnamefont {Kastner}}, \bibinfo {author} {\bibfnamefont
  {T.}~\bibnamefont {Frank}}, \bibinfo {author} {\bibfnamefont
  {J.}~\bibnamefont {Fabian}}, \bibinfo {author} {\bibfnamefont {M.~S.}\
  \bibnamefont {Vitiello}}, \bibinfo {author} {\bibfnamefont {T.~L.}\
  \bibnamefont {Cocker}}, \ and\ \bibinfo {author} {\bibfnamefont
  {R.}~\bibnamefont {Huber}},\ }\href {\doibase 10.1038/nnano.2016.261}
  {\bibfield  {journal} {\bibinfo  {journal} {Nat. Nanotechnol.}\ }\textbf
  {\bibinfo {volume} {12}},\ \bibinfo {pages} {207} (\bibinfo {year}
  {2016})}\BibitemShut {NoStop}%
\bibitem [{\citenamefont {Amenabar}\ \emph {et~al.}(2017)\citenamefont
  {Amenabar}, \citenamefont {Poly}, \citenamefont {Goikoetxea}, \citenamefont
  {Nuansing}, \citenamefont {Lasch},\ and\ \citenamefont
  {Hillenbrand}}]{Amenabar2017}%
  \BibitemOpen
  \bibfield  {author} {\bibinfo {author} {\bibfnamefont {I.}~\bibnamefont
  {Amenabar}}, \bibinfo {author} {\bibfnamefont {S.}~\bibnamefont {Poly}},
  \bibinfo {author} {\bibfnamefont {M.}~\bibnamefont {Goikoetxea}}, \bibinfo
  {author} {\bibfnamefont {W.}~\bibnamefont {Nuansing}}, \bibinfo {author}
  {\bibfnamefont {P.}~\bibnamefont {Lasch}}, \ and\ \bibinfo {author}
  {\bibfnamefont {R.}~\bibnamefont {Hillenbrand}},\ }\href {\doibase
  10.1038/ncomms14402} {\bibfield  {journal} {\bibinfo  {journal} {Nat.
  Commun.}\ }\textbf {\bibinfo {volume} {8}},\ \bibinfo {pages} {14402}
  (\bibinfo {year} {2017})}\BibitemShut {NoStop}%
\bibitem [{\citenamefont {Knoll}\ and\ \citenamefont
  {Keilmann}(2000)}]{Knoll2000}%
  \BibitemOpen
  \bibfield  {author} {\bibinfo {author} {\bibfnamefont {B.}~\bibnamefont
  {Knoll}}\ and\ \bibinfo {author} {\bibfnamefont {F.}~\bibnamefont
  {Keilmann}},\ }\href {\doibase 10.1016/S0030-4018(00)00826-9} {\bibfield
  {journal} {\bibinfo  {journal} {Opt. Commun.}\ }\textbf {\bibinfo {volume}
  {182}},\ \bibinfo {pages} {321} (\bibinfo {year} {2000})}\BibitemShut
  {NoStop}%
\bibitem [{\citenamefont {Ni}\ \emph {et~al.}(2016)\citenamefont {Ni},
  \citenamefont {Wang}, \citenamefont {Goldflam}, \citenamefont {Wagner},
  \citenamefont {Fei}, \citenamefont {McLeod}, \citenamefont {Liu},
  \citenamefont {Keilmann}, \citenamefont {{\"{O}}zyilmaz}, \citenamefont
  {{Castro Neto}}, \citenamefont {Hone}, \citenamefont {Fogler},\ and\
  \citenamefont {Basov}}]{Ni2016}%
  \BibitemOpen
  \bibfield  {author} {\bibinfo {author} {\bibfnamefont {G.~X.}\ \bibnamefont
  {Ni}}, \bibinfo {author} {\bibfnamefont {L.}~\bibnamefont {Wang}}, \bibinfo
  {author} {\bibfnamefont {M.~D.}\ \bibnamefont {Goldflam}}, \bibinfo {author}
  {\bibfnamefont {M.}~\bibnamefont {Wagner}}, \bibinfo {author} {\bibfnamefont
  {Z.}~\bibnamefont {Fei}}, \bibinfo {author} {\bibfnamefont {A.~S.}\
  \bibnamefont {McLeod}}, \bibinfo {author} {\bibfnamefont {M.~K.}\
  \bibnamefont {Liu}}, \bibinfo {author} {\bibfnamefont {F.}~\bibnamefont
  {Keilmann}}, \bibinfo {author} {\bibfnamefont {B.}~\bibnamefont
  {{\"{O}}zyilmaz}}, \bibinfo {author} {\bibfnamefont {A.~H.}\ \bibnamefont
  {{Castro Neto}}}, \bibinfo {author} {\bibfnamefont {J.}~\bibnamefont {Hone}},
  \bibinfo {author} {\bibfnamefont {M.~M.}\ \bibnamefont {Fogler}}, \ and\
  \bibinfo {author} {\bibfnamefont {D.~N.}\ \bibnamefont {Basov}},\ }\href
  {http://www.nature.com/doifinder/10.1038/nphoton.2016.45} {\bibfield
  {journal} {\bibinfo  {journal} {Nat. Photonics}\ }\textbf {\bibinfo {volume}
  {10}},\ \bibinfo {pages} {244} (\bibinfo {year} {2016})}\BibitemShut
  {NoStop}%
\bibitem [{\citenamefont {Jiang}\ \emph {et~al.}(2016)\citenamefont {Jiang},
  \citenamefont {Shi}, \citenamefont {Zeng}, \citenamefont {Wang},
  \citenamefont {Kang}, \citenamefont {Joshi}, \citenamefont {Jin},
  \citenamefont {Ju}, \citenamefont {Kim}, \citenamefont {Lyu}, \citenamefont
  {Shen}, \citenamefont {Crommie}, \citenamefont {Gao},\ and\ \citenamefont
  {Wang}}]{Jiang2016}%
  \BibitemOpen
  \bibfield  {author} {\bibinfo {author} {\bibfnamefont {L.}~\bibnamefont
  {Jiang}}, \bibinfo {author} {\bibfnamefont {Z.}~\bibnamefont {Shi}}, \bibinfo
  {author} {\bibfnamefont {B.}~\bibnamefont {Zeng}}, \bibinfo {author}
  {\bibfnamefont {S.}~\bibnamefont {Wang}}, \bibinfo {author} {\bibfnamefont
  {J.-H.}\ \bibnamefont {Kang}}, \bibinfo {author} {\bibfnamefont
  {T.}~\bibnamefont {Joshi}}, \bibinfo {author} {\bibfnamefont
  {C.}~\bibnamefont {Jin}}, \bibinfo {author} {\bibfnamefont {L.}~\bibnamefont
  {Ju}}, \bibinfo {author} {\bibfnamefont {J.}~\bibnamefont {Kim}}, \bibinfo
  {author} {\bibfnamefont {T.}~\bibnamefont {Lyu}}, \bibinfo {author}
  {\bibfnamefont {Y.-R.}\ \bibnamefont {Shen}}, \bibinfo {author}
  {\bibfnamefont {M.}~\bibnamefont {Crommie}}, \bibinfo {author} {\bibfnamefont
  {H.-J.}\ \bibnamefont {Gao}}, \ and\ \bibinfo {author} {\bibfnamefont
  {F.}~\bibnamefont {Wang}},\ }\href {\doibase 10.1038/nmat4653} {\bibfield
  {journal} {\bibinfo  {journal} {Nat. Mater.}\ }\textbf {\bibinfo {volume}
  {15}},\ \bibinfo {pages} {1} (\bibinfo {year} {2016})}\BibitemShut {NoStop}%
\bibitem [{\citenamefont {Alonso-Gonz{\'{a}}lez}\ \emph
  {et~al.}(2016)\citenamefont {Alonso-Gonz{\'{a}}lez}, \citenamefont {Nikitin},
  \citenamefont {Gao}, \citenamefont {Woessner}, \citenamefont {Lundeberg},
  \citenamefont {Principi}, \citenamefont {Forcellini}, \citenamefont {Yan},
  \citenamefont {V{\'{e}}lez}, \citenamefont {Huber}, \citenamefont {Watanabe},
  \citenamefont {Taniguchi}, \citenamefont {Casanova}, \citenamefont {Hueso},
  \citenamefont {Polini}, \citenamefont {Hone}, \citenamefont {Koppens},\ and\
  \citenamefont {Hillenbrand}}]{Alonso-Gonzalez2016}%
  \BibitemOpen
  \bibfield  {author} {\bibinfo {author} {\bibfnamefont {P.}~\bibnamefont
  {Alonso-Gonz{\'{a}}lez}}, \bibinfo {author} {\bibfnamefont {A.~Y.}\
  \bibnamefont {Nikitin}}, \bibinfo {author} {\bibfnamefont {Y.}~\bibnamefont
  {Gao}}, \bibinfo {author} {\bibfnamefont {A.}~\bibnamefont {Woessner}},
  \bibinfo {author} {\bibfnamefont {M.~B.}\ \bibnamefont {Lundeberg}}, \bibinfo
  {author} {\bibfnamefont {A.}~\bibnamefont {Principi}}, \bibinfo {author}
  {\bibfnamefont {N.}~\bibnamefont {Forcellini}}, \bibinfo {author}
  {\bibfnamefont {W.}~\bibnamefont {Yan}}, \bibinfo {author} {\bibfnamefont
  {S.}~\bibnamefont {V{\'{e}}lez}}, \bibinfo {author} {\bibfnamefont {A.~J.}\
  \bibnamefont {Huber}}, \bibinfo {author} {\bibfnamefont {K.}~\bibnamefont
  {Watanabe}}, \bibinfo {author} {\bibfnamefont {T.}~\bibnamefont {Taniguchi}},
  \bibinfo {author} {\bibfnamefont {F.}~\bibnamefont {Casanova}}, \bibinfo
  {author} {\bibfnamefont {L.~E.}\ \bibnamefont {Hueso}}, \bibinfo {author}
  {\bibfnamefont {M.}~\bibnamefont {Polini}}, \bibinfo {author} {\bibfnamefont
  {J.}~\bibnamefont {Hone}}, \bibinfo {author} {\bibfnamefont {F.~H.~L.}\
  \bibnamefont {Koppens}}, \ and\ \bibinfo {author} {\bibfnamefont
  {R.}~\bibnamefont {Hillenbrand}},\ }\href
  {http://arxiv.org/abs/1601.05753%5Cnhttp://www.nature.com/doifinder/10.1038/nnano.2016.185}
  {\bibfield  {journal} {\bibinfo  {journal} {Nat. Nanotechnol.}\ }\textbf
  {\bibinfo {volume} {12}},\ \bibinfo {pages} {1} (\bibinfo {year}
  {2016})}\BibitemShut {NoStop}%
\bibitem [{\citenamefont {Patoka}\ \emph {et~al.}(2016)\citenamefont {Patoka},
  \citenamefont {Ulrich}, \citenamefont {Nguyen}, \citenamefont {Bartels},
  \citenamefont {Dowben}, \citenamefont {Turkowski}, \citenamefont {Rahman},
  \citenamefont {Hermann}, \citenamefont {Kastner}, \citenamefont {Hoehl},
  \citenamefont {Ulm},\ and\ \citenamefont {Ruhl}}]{Patoka2016}%
  \BibitemOpen
  \bibfield  {author} {\bibinfo {author} {\bibfnamefont {P.}~\bibnamefont
  {Patoka}}, \bibinfo {author} {\bibfnamefont {G.}~\bibnamefont {Ulrich}},
  \bibinfo {author} {\bibfnamefont {A.~E.}\ \bibnamefont {Nguyen}}, \bibinfo
  {author} {\bibfnamefont {L.}~\bibnamefont {Bartels}}, \bibinfo {author}
  {\bibfnamefont {P.~A.}\ \bibnamefont {Dowben}}, \bibinfo {author}
  {\bibfnamefont {V.}~\bibnamefont {Turkowski}}, \bibinfo {author}
  {\bibfnamefont {T.~S.}\ \bibnamefont {Rahman}}, \bibinfo {author}
  {\bibfnamefont {P.}~\bibnamefont {Hermann}}, \bibinfo {author} {\bibfnamefont
  {B.}~\bibnamefont {Kastner}}, \bibinfo {author} {\bibfnamefont
  {A.}~\bibnamefont {Hoehl}}, \bibinfo {author} {\bibfnamefont
  {G.}~\bibnamefont {Ulm}}, \ and\ \bibinfo {author} {\bibfnamefont
  {E.}~\bibnamefont {Ruhl}},\ }\href {\doibase 10.1364/oe.24.001154} {\bibfield
   {journal} {\bibinfo  {journal} {Opt. Express}\ }\textbf {\bibinfo {volume}
  {24}},\ \bibinfo {pages} {1154} (\bibinfo {year} {2016})}\BibitemShut
  {NoStop}%
\bibitem [{\citenamefont {Taubner}\ \emph {et~al.}(2006)\citenamefont
  {Taubner}, \citenamefont {Korobkin}, \citenamefont {Urzhumov}, \citenamefont
  {Shvets},\ and\ \citenamefont {Hillenbrand}}]{Taubner2006}%
  \BibitemOpen
  \bibfield  {author} {\bibinfo {author} {\bibfnamefont {T.}~\bibnamefont
  {Taubner}}, \bibinfo {author} {\bibfnamefont {D.}~\bibnamefont {Korobkin}},
  \bibinfo {author} {\bibfnamefont {Y.}~\bibnamefont {Urzhumov}}, \bibinfo
  {author} {\bibfnamefont {G.}~\bibnamefont {Shvets}}, \ and\ \bibinfo {author}
  {\bibfnamefont {R.}~\bibnamefont {Hillenbrand}},\ }\href {\doibase
  10.1126/science.1131025} {\bibfield  {journal} {\bibinfo  {journal}
  {Science}\ }\textbf {\bibinfo {volume} {313}},\ \bibinfo {pages} {1595}
  (\bibinfo {year} {2006})}\BibitemShut {NoStop}%
\bibitem [{\citenamefont {Kehr}\ \emph {et~al.}(2011)\citenamefont {Kehr},
  \citenamefont {Liu}, \citenamefont {Martin}, \citenamefont {Yu},
  \citenamefont {Gajek}, \citenamefont {Yang}, \citenamefont {Yang},
  \citenamefont {Wenzel}, \citenamefont {Jacob}, \citenamefont {von Ribbeck},
  \citenamefont {Helm}, \citenamefont {Zhang}, \citenamefont {Eng},\ and\
  \citenamefont {Ramesh}}]{Kehr2011}%
  \BibitemOpen
  \bibfield  {author} {\bibinfo {author} {\bibfnamefont {S.~C.}\ \bibnamefont
  {Kehr}}, \bibinfo {author} {\bibfnamefont {Y.~M.}\ \bibnamefont {Liu}},
  \bibinfo {author} {\bibfnamefont {L.~W.}\ \bibnamefont {Martin}}, \bibinfo
  {author} {\bibfnamefont {P.}~\bibnamefont {Yu}}, \bibinfo {author}
  {\bibfnamefont {M.}~\bibnamefont {Gajek}}, \bibinfo {author} {\bibfnamefont
  {S.-Y.}\ \bibnamefont {Yang}}, \bibinfo {author} {\bibfnamefont {C.-H.}\
  \bibnamefont {Yang}}, \bibinfo {author} {\bibfnamefont {M.~T.}\ \bibnamefont
  {Wenzel}}, \bibinfo {author} {\bibfnamefont {R.}~\bibnamefont {Jacob}},
  \bibinfo {author} {\bibfnamefont {H.-G.}\ \bibnamefont {von Ribbeck}},
  \bibinfo {author} {\bibfnamefont {M.}~\bibnamefont {Helm}}, \bibinfo {author}
  {\bibfnamefont {X.}~\bibnamefont {Zhang}}, \bibinfo {author} {\bibfnamefont
  {L.~M.}\ \bibnamefont {Eng}}, \ and\ \bibinfo {author} {\bibfnamefont
  {R.}~\bibnamefont {Ramesh}},\ }\href
  {http://www.pubmedcentral.nih.gov/articlerender.fcgi?artid=3072079&tool=pmcentrez&rendertype=abstract}
  {\bibfield  {journal} {\bibinfo  {journal} {Nat. Commun.}\ }\textbf {\bibinfo
  {volume} {2}},\ \bibinfo {pages} {249} (\bibinfo {year} {2011})}\BibitemShut
  {NoStop}%
\bibitem [{\citenamefont {Fehrenbacher}\ \emph {et~al.}(2015)\citenamefont
  {Fehrenbacher}, \citenamefont {Winnerl}, \citenamefont {Schneider},
  \citenamefont {D{\"{o}}ring}, \citenamefont {Kehr}, \citenamefont {Eng},
  \citenamefont {Huo}, \citenamefont {Schmidt}, \citenamefont {Yao},
  \citenamefont {Liu},\ and\ \citenamefont {Helm}}]{Fehrenbacher2015}%
  \BibitemOpen
  \bibfield  {author} {\bibinfo {author} {\bibfnamefont {M.}~\bibnamefont
  {Fehrenbacher}}, \bibinfo {author} {\bibfnamefont {S.}~\bibnamefont
  {Winnerl}}, \bibinfo {author} {\bibfnamefont {H.}~\bibnamefont {Schneider}},
  \bibinfo {author} {\bibfnamefont {J.}~\bibnamefont {D{\"{o}}ring}}, \bibinfo
  {author} {\bibfnamefont {S.~C.}\ \bibnamefont {Kehr}}, \bibinfo {author}
  {\bibfnamefont {L.~M.}\ \bibnamefont {Eng}}, \bibinfo {author} {\bibfnamefont
  {Y.}~\bibnamefont {Huo}}, \bibinfo {author} {\bibfnamefont {O.~G.}\
  \bibnamefont {Schmidt}}, \bibinfo {author} {\bibfnamefont {K.}~\bibnamefont
  {Yao}}, \bibinfo {author} {\bibfnamefont {Y.}~\bibnamefont {Liu}}, \ and\
  \bibinfo {author} {\bibfnamefont {M.}~\bibnamefont {Helm}},\ }\href {\doibase
  10.1021/nl503996q} {\bibfield  {journal} {\bibinfo  {journal} {Nano Lett.}\
  }\textbf {\bibinfo {volume} {15}},\ \bibinfo {pages} {1057} (\bibinfo {year}
  {2015})}\BibitemShut {NoStop}%
\bibitem [{\citenamefont {Westermeier}\ \emph {et~al.}(2014)\citenamefont
  {Westermeier}, \citenamefont {Cernescu}, \citenamefont {Amarie},
  \citenamefont {Liewald}, \citenamefont {Keilmann},\ and\ \citenamefont
  {Nickel}}]{Westermeier2014}%
  \BibitemOpen
  \bibfield  {author} {\bibinfo {author} {\bibfnamefont {C.}~\bibnamefont
  {Westermeier}}, \bibinfo {author} {\bibfnamefont {A.}~\bibnamefont
  {Cernescu}}, \bibinfo {author} {\bibfnamefont {S.}~\bibnamefont {Amarie}},
  \bibinfo {author} {\bibfnamefont {C.}~\bibnamefont {Liewald}}, \bibinfo
  {author} {\bibfnamefont {F.}~\bibnamefont {Keilmann}}, \ and\ \bibinfo
  {author} {\bibfnamefont {B.}~\bibnamefont {Nickel}},\ }\href {\doibase
  10.1038/ncomms5101} {\bibfield  {journal} {\bibinfo  {journal} {Nat.
  Commun.}\ }\textbf {\bibinfo {volume} {5}},\ \bibinfo {pages} {4101}
  (\bibinfo {year} {2014})}\BibitemShut {NoStop}%
\bibitem [{\citenamefont {Ozyuzer}\ \emph {et~al.}(2007)\citenamefont
  {Ozyuzer}, \citenamefont {Koshelev}, \citenamefont {Kurter}, \citenamefont
  {Gopalsami}, \citenamefont {Li}, \citenamefont {Tachiki}, \citenamefont
  {Kadowaki}, \citenamefont {Yamamoto}, \citenamefont {Minami}, \citenamefont
  {Yamaguchi}, \citenamefont {Tachiki}, \citenamefont {Gray}, \citenamefont
  {Kwok},\ and\ \citenamefont {Welp}}]{Yamamoto2007}%
  \BibitemOpen
  \bibfield  {author} {\bibinfo {author} {\bibfnamefont {L.}~\bibnamefont
  {Ozyuzer}}, \bibinfo {author} {\bibfnamefont {a.~E.}\ \bibnamefont
  {Koshelev}}, \bibinfo {author} {\bibfnamefont {C.}~\bibnamefont {Kurter}},
  \bibinfo {author} {\bibfnamefont {N.}~\bibnamefont {Gopalsami}}, \bibinfo
  {author} {\bibfnamefont {Q.}~\bibnamefont {Li}}, \bibinfo {author}
  {\bibfnamefont {M.}~\bibnamefont {Tachiki}}, \bibinfo {author} {\bibfnamefont
  {K.}~\bibnamefont {Kadowaki}}, \bibinfo {author} {\bibfnamefont
  {T.}~\bibnamefont {Yamamoto}}, \bibinfo {author} {\bibfnamefont
  {H.}~\bibnamefont {Minami}}, \bibinfo {author} {\bibfnamefont
  {H.}~\bibnamefont {Yamaguchi}}, \bibinfo {author} {\bibfnamefont
  {T.}~\bibnamefont {Tachiki}}, \bibinfo {author} {\bibfnamefont {K.~E.}\
  \bibnamefont {Gray}}, \bibinfo {author} {\bibfnamefont {W.-K.}\ \bibnamefont
  {Kwok}}, \ and\ \bibinfo {author} {\bibfnamefont {U.}~\bibnamefont {Welp}},\
  }\href {\doibase 10.1126/science.1149802} {\bibfield  {journal} {\bibinfo
  {journal} {Science}\ }\textbf {\bibinfo {volume} {318}},\ \bibinfo {pages}
  {1291} (\bibinfo {year} {2007})}\BibitemShut {NoStop}%
\bibitem [{\citenamefont {Ahmad}\ \emph {et~al.}(2015)\citenamefont {Ahmad},
  \citenamefont {Min}, \citenamefont {Seo}, \citenamefont {Choi}, \citenamefont
  {Kimura}, \citenamefont {Seo},\ and\ \citenamefont {Kwon}}]{Ahmad2015}%
  \BibitemOpen
  \bibfield  {author} {\bibinfo {author} {\bibfnamefont {D.}~\bibnamefont
  {Ahmad}}, \bibinfo {author} {\bibfnamefont {B.~H.}\ \bibnamefont {Min}},
  \bibinfo {author} {\bibfnamefont {Y.~I.}\ \bibnamefont {Seo}}, \bibinfo
  {author} {\bibfnamefont {W.~J.}\ \bibnamefont {Choi}}, \bibinfo {author}
  {\bibfnamefont {S.-I.}\ \bibnamefont {Kimura}}, \bibinfo {author}
  {\bibfnamefont {J.}~\bibnamefont {Seo}}, \ and\ \bibinfo {author}
  {\bibfnamefont {Y.~S.}\ \bibnamefont {Kwon}},\ }\href {\doibase
  10.1088/0953-2048/28/7/075002} {\bibfield  {journal} {\bibinfo  {journal}
  {Supercond. Sci. Technol.}\ }\textbf {\bibinfo {volume} {28}},\ \bibinfo
  {pages} {075002} (\bibinfo {year} {2015})}\BibitemShut {NoStop}%
\bibitem [{\citenamefont {D{\"{o}}ring}\ \emph {et~al.}(2014)\citenamefont
  {D{\"{o}}ring}, \citenamefont {non Ribbeck}, \citenamefont {Fehrenbacher},
  \citenamefont {Kehr},\ and\ \citenamefont {Eng}}]{Doring2014}%
  \BibitemOpen
  \bibfield  {author} {\bibinfo {author} {\bibfnamefont {J.}~\bibnamefont
  {D{\"{o}}ring}}, \bibinfo {author} {\bibfnamefont {H.-G.}\ \bibnamefont {non
  Ribbeck}}, \bibinfo {author} {\bibfnamefont {M.}~\bibnamefont
  {Fehrenbacher}}, \bibinfo {author} {\bibfnamefont {S.~C.}\ \bibnamefont
  {Kehr}}, \ and\ \bibinfo {author} {\bibfnamefont {L.~M.}\ \bibnamefont
  {Eng}},\ }\href@noop {} {\bibfield  {journal} {\bibinfo  {journal} {Appl.
  Phys. Lett.}\ }\textbf {\bibinfo {volume} {105}},\ \bibinfo {pages} {053109}
  (\bibinfo {year} {2014})}\BibitemShut {NoStop}%
\bibitem [{\citenamefont {McLeod}\ \emph {et~al.}(2016)\citenamefont {McLeod},
  \citenamefont {van Heumen}, \citenamefont {Ramirez}, \citenamefont {Wang},
  \citenamefont {Saerbeck}, \citenamefont {Guenon}, \citenamefont {Goldflam},
  \citenamefont {Anderegg}, \citenamefont {Kelly}, \citenamefont {Mueller},
  \citenamefont {Liu}, \citenamefont {Schuller},\ and\ \citenamefont
  {Basov}}]{McLeod2016}%
  \BibitemOpen
  \bibfield  {author} {\bibinfo {author} {\bibfnamefont {A.~S.}\ \bibnamefont
  {McLeod}}, \bibinfo {author} {\bibfnamefont {E.}~\bibnamefont {van Heumen}},
  \bibinfo {author} {\bibfnamefont {J.~G.}\ \bibnamefont {Ramirez}}, \bibinfo
  {author} {\bibfnamefont {S.}~\bibnamefont {Wang}}, \bibinfo {author}
  {\bibfnamefont {T.}~\bibnamefont {Saerbeck}}, \bibinfo {author}
  {\bibfnamefont {S.}~\bibnamefont {Guenon}}, \bibinfo {author} {\bibfnamefont
  {M.}~\bibnamefont {Goldflam}}, \bibinfo {author} {\bibfnamefont
  {L.}~\bibnamefont {Anderegg}}, \bibinfo {author} {\bibfnamefont
  {P.}~\bibnamefont {Kelly}}, \bibinfo {author} {\bibfnamefont
  {A.}~\bibnamefont {Mueller}}, \bibinfo {author} {\bibfnamefont {M.~K.}\
  \bibnamefont {Liu}}, \bibinfo {author} {\bibfnamefont {I.~K.}\ \bibnamefont
  {Schuller}}, \ and\ \bibinfo {author} {\bibfnamefont {D.~N.}\ \bibnamefont
  {Basov}},\ }\href {\doibase 10.1038/nphys3882} {\bibfield  {journal}
  {\bibinfo  {journal} {Nat. Phys.}\ }\textbf {\bibinfo {volume} {13}},\
  \bibinfo {pages} {80} (\bibinfo {year} {2016})}\BibitemShut {NoStop}%
\bibitem [{\citenamefont {Pajot}(2010)}]{Pajot2010}%
  \BibitemOpen
  \bibfield  {author} {\bibinfo {author} {\bibfnamefont {B.}~\bibnamefont
  {Pajot}},\ }\href {\doibase 10.1007/b135694} {\emph {\bibinfo {title}
  {{Optical Absorption of Impurities and Defects in Semiconducting
  Crystals}}}},\ \bibinfo {series} {Springer Series in Solid-State Sciences},
  Vol.\ \bibinfo {volume} {158}\ (\bibinfo  {publisher} {Springer Berlin
  Heidelberg},\ \bibinfo {address} {Berlin, Heidelberg},\ \bibinfo {year}
  {2010})\BibitemShut {NoStop}%
\bibitem [{\citenamefont {Moldovan-Doyen}\ \emph {et~al.}(2011)\citenamefont
  {Moldovan-Doyen}, \citenamefont {Xu}, \citenamefont {Greusard}, \citenamefont
  {Sevin}, \citenamefont {Strupiechonski}, \citenamefont {Beaudoin},
  \citenamefont {Sagnes}, \citenamefont {Khanna}, \citenamefont {Linfield},
  \citenamefont {Davies}, \citenamefont {Colombelli},\ and\ \citenamefont {{De
  Wilde}}}]{Moldovan-Doyen2011}%
  \BibitemOpen
  \bibfield  {author} {\bibinfo {author} {\bibfnamefont {I.~C.}\ \bibnamefont
  {Moldovan-Doyen}}, \bibinfo {author} {\bibfnamefont {G.}~\bibnamefont {Xu}},
  \bibinfo {author} {\bibfnamefont {L.}~\bibnamefont {Greusard}}, \bibinfo
  {author} {\bibfnamefont {G.}~\bibnamefont {Sevin}}, \bibinfo {author}
  {\bibfnamefont {E.}~\bibnamefont {Strupiechonski}}, \bibinfo {author}
  {\bibfnamefont {G.}~\bibnamefont {Beaudoin}}, \bibinfo {author}
  {\bibfnamefont {I.}~\bibnamefont {Sagnes}}, \bibinfo {author} {\bibfnamefont
  {S.~P.}\ \bibnamefont {Khanna}}, \bibinfo {author} {\bibfnamefont {E.~H.}\
  \bibnamefont {Linfield}}, \bibinfo {author} {\bibfnamefont {A.~G.}\
  \bibnamefont {Davies}}, \bibinfo {author} {\bibfnamefont {R.}~\bibnamefont
  {Colombelli}}, \ and\ \bibinfo {author} {\bibfnamefont {Y.}~\bibnamefont {{De
  Wilde}}},\ }\href@noop {} {\bibfield  {journal} {\bibinfo  {journal} {Appl.
  Phys. Lett.}\ }\textbf {\bibinfo {volume} {98}},\ \bibinfo {pages} {231112}
  (\bibinfo {year} {2011})}\BibitemShut {NoStop}%
\bibitem [{\citenamefont {Yang}\ \emph {et~al.}(2013)\citenamefont {Yang},
  \citenamefont {Hebestreit}, \citenamefont {Josberger},\ and\ \citenamefont
  {Raschke}}]{Yang2013}%
  \BibitemOpen
  \bibfield  {author} {\bibinfo {author} {\bibfnamefont {H.~U.}\ \bibnamefont
  {Yang}}, \bibinfo {author} {\bibfnamefont {E.}~\bibnamefont {Hebestreit}},
  \bibinfo {author} {\bibfnamefont {E.~E.}\ \bibnamefont {Josberger}}, \ and\
  \bibinfo {author} {\bibfnamefont {M.~B.}\ \bibnamefont {Raschke}},\ }\href
  {\doibase 10.1063/1.4789428} {\bibfield  {journal} {\bibinfo  {journal} {Rev.
  Sci. Instrum.}\ }\textbf {\bibinfo {volume} {84}},\ \bibinfo {pages} {023701}
  (\bibinfo {year} {2013})}\BibitemShut {NoStop}%
\bibitem [{\citenamefont {Milde}\ \emph {et~al.}(2013)\citenamefont {Milde},
  \citenamefont {K{\"{o}}hler}, \citenamefont {Seidel}, \citenamefont {Eng},
  \citenamefont {Bauer}, \citenamefont {Chacon}, \citenamefont {Kindervater},
  \citenamefont {M{\"{u}}hlbauer}, \citenamefont {Pfleiderer}, \citenamefont
  {Buhrandt}, \citenamefont {Sch{\"{u}}tte},\ and\ \citenamefont
  {Rosch}}]{Milde2013}%
  \BibitemOpen
  \bibfield  {author} {\bibinfo {author} {\bibfnamefont {P.}~\bibnamefont
  {Milde}}, \bibinfo {author} {\bibfnamefont {D.}~\bibnamefont {K{\"{o}}hler}},
  \bibinfo {author} {\bibfnamefont {J.}~\bibnamefont {Seidel}}, \bibinfo
  {author} {\bibfnamefont {L.~M.}\ \bibnamefont {Eng}}, \bibinfo {author}
  {\bibfnamefont {A.}~\bibnamefont {Bauer}}, \bibinfo {author} {\bibfnamefont
  {A.}~\bibnamefont {Chacon}}, \bibinfo {author} {\bibfnamefont
  {J.}~\bibnamefont {Kindervater}}, \bibinfo {author} {\bibfnamefont
  {S.}~\bibnamefont {M{\"{u}}hlbauer}}, \bibinfo {author} {\bibfnamefont
  {C.}~\bibnamefont {Pfleiderer}}, \bibinfo {author} {\bibfnamefont
  {S.}~\bibnamefont {Buhrandt}}, \bibinfo {author} {\bibfnamefont
  {C.}~\bibnamefont {Sch{\"{u}}tte}}, \ and\ \bibinfo {author} {\bibfnamefont
  {A.}~\bibnamefont {Rosch}},\ }\href {\doibase 10.1126/science.1234657}
  {\bibfield  {journal} {\bibinfo  {journal} {Science}\ }\textbf {\bibinfo
  {volume} {340}},\ \bibinfo {pages} {1076} (\bibinfo {year}
  {2013})}\BibitemShut {NoStop}%
\bibitem [{\citenamefont {Wang}\ \emph {et~al.}(2015)\citenamefont {Wang},
  \citenamefont {Ruff}, \citenamefont {Schmidt}, \citenamefont {Tsurkan},
  \citenamefont {K{\'{e}}zsm{\'{a}}rki}, \citenamefont {Lunkenheimer},\ and\
  \citenamefont {Loidl}}]{Wang2015}%
  \BibitemOpen
  \bibfield  {author} {\bibinfo {author} {\bibfnamefont {Z.}~\bibnamefont
  {Wang}}, \bibinfo {author} {\bibfnamefont {E.}~\bibnamefont {Ruff}}, \bibinfo
  {author} {\bibfnamefont {M.}~\bibnamefont {Schmidt}}, \bibinfo {author}
  {\bibfnamefont {V.}~\bibnamefont {Tsurkan}}, \bibinfo {author} {\bibfnamefont
  {I.}~\bibnamefont {K{\'{e}}zsm{\'{a}}rki}}, \bibinfo {author} {\bibfnamefont
  {P.}~\bibnamefont {Lunkenheimer}}, \ and\ \bibinfo {author} {\bibfnamefont
  {A.}~\bibnamefont {Loidl}},\ }\href {\doibase 10.1103/PhysRevLett.115.207601}
  {\bibfield  {journal} {\bibinfo  {journal} {Phys. Rev. Lett.}\ }\textbf
  {\bibinfo {volume} {115}},\ \bibinfo {pages} {1} (\bibinfo {year}
  {2015})}\BibitemShut {NoStop}%
\bibitem [{\citenamefont {K{\'{e}}zsm{\'{a}}rki}\ \emph
  {et~al.}(2015)\citenamefont {K{\'{e}}zsm{\'{a}}rki}, \citenamefont
  {Bord{\'{a}}cs}, \citenamefont {Milde}, \citenamefont {Neuber}, \citenamefont
  {Eng}, \citenamefont {White}, \citenamefont {R{\o}nnow}, \citenamefont
  {Dewhurst}, \citenamefont {Mochizuki}, \citenamefont {Yanai}, \citenamefont
  {Nakamura}, \citenamefont {Ehlers}, \citenamefont {Tsurkan},\ and\
  \citenamefont {Loidl}}]{Kezsmarki2015}%
  \BibitemOpen
  \bibfield  {author} {\bibinfo {author} {\bibfnamefont {I.}~\bibnamefont
  {K{\'{e}}zsm{\'{a}}rki}}, \bibinfo {author} {\bibfnamefont {S.}~\bibnamefont
  {Bord{\'{a}}cs}}, \bibinfo {author} {\bibfnamefont {P.}~\bibnamefont
  {Milde}}, \bibinfo {author} {\bibfnamefont {E.}~\bibnamefont {Neuber}},
  \bibinfo {author} {\bibfnamefont {L.~M.}\ \bibnamefont {Eng}}, \bibinfo
  {author} {\bibfnamefont {J.~S.}\ \bibnamefont {White}}, \bibinfo {author}
  {\bibfnamefont {H.~M.}\ \bibnamefont {R{\o}nnow}}, \bibinfo {author}
  {\bibfnamefont {C.~D.}\ \bibnamefont {Dewhurst}}, \bibinfo {author}
  {\bibfnamefont {M.}~\bibnamefont {Mochizuki}}, \bibinfo {author}
  {\bibfnamefont {K.}~\bibnamefont {Yanai}}, \bibinfo {author} {\bibfnamefont
  {H.}~\bibnamefont {Nakamura}}, \bibinfo {author} {\bibfnamefont
  {D.}~\bibnamefont {Ehlers}}, \bibinfo {author} {\bibfnamefont
  {V.}~\bibnamefont {Tsurkan}}, \ and\ \bibinfo {author} {\bibfnamefont
  {A.}~\bibnamefont {Loidl}},\ }\href {http://dx.doi.org/10.1038/nmat4402}
  {\bibfield  {journal} {\bibinfo  {journal} {Nat. Mater.}\ }\textbf {\bibinfo
  {volume} {14}},\ \bibinfo {pages} {1116} (\bibinfo {year}
  {2015})}\BibitemShut {NoStop}%
\bibitem [{\citenamefont {Butykai}\ \emph {et~al.}(2017)\citenamefont
  {Butykai}, \citenamefont {Bord{\'{a}}cs}, \citenamefont
  {K{\'{e}}zsm{\'{a}}rki}, \citenamefont {Tsurkan}, \citenamefont {Loidl},
  \citenamefont {D{\"{o}}ring}, \citenamefont {Milde}, \citenamefont {Kehr},\
  and\ \citenamefont {Eng}}]{Butykai2017}%
  \BibitemOpen
  \bibfield  {author} {\bibinfo {author} {\bibfnamefont {{\'{A}}.}~\bibnamefont
  {Butykai}}, \bibinfo {author} {\bibfnamefont {S.}~\bibnamefont
  {Bord{\'{a}}cs}}, \bibinfo {author} {\bibfnamefont {I.}~\bibnamefont
  {K{\'{e}}zsm{\'{a}}rki}}, \bibinfo {author} {\bibfnamefont {V.}~\bibnamefont
  {Tsurkan}}, \bibinfo {author} {\bibfnamefont {A.}~\bibnamefont {Loidl}},
  \bibinfo {author} {\bibfnamefont {J.}~\bibnamefont {D{\"{o}}ring}}, \bibinfo
  {author} {\bibfnamefont {P.}~\bibnamefont {Milde}}, \bibinfo {author}
  {\bibfnamefont {S.~C.}\ \bibnamefont {Kehr}}, \ and\ \bibinfo {author}
  {\bibfnamefont {L.~M.}\ \bibnamefont {Eng}},\ }\href
  {http://arxiv.org/abs/1702.07245} {\bibfield  {journal} {\bibinfo  {journal}
  {Sci. Rep.}\ }\textbf {\bibinfo {volume} {7}},\ \bibinfo {pages} {44663}
  (\bibinfo {year} {2017})}\BibitemShut {NoStop}%
\bibitem [{\citenamefont {D{\"{o}}ring}, \citenamefont {Eng},\ and\
  \citenamefont {Kehr}(2016)}]{Doring2016a}%
  \BibitemOpen
  \bibfield  {author} {\bibinfo {author} {\bibfnamefont {J.}~\bibnamefont
  {D{\"{o}}ring}}, \bibinfo {author} {\bibfnamefont {L.~M.}\ \bibnamefont
  {Eng}}, \ and\ \bibinfo {author} {\bibfnamefont {S.~C.}\ \bibnamefont
  {Kehr}},\ }\href@noop {} {\bibfield  {journal} {\bibinfo  {journal} {J. Appl.
  Phys.}\ }\textbf {\bibinfo {volume} {120}} (\bibinfo {year}
  {2016})}\BibitemShut {NoStop}%
\bibitem [{\citenamefont {Schneider}, \citenamefont {Grafstr{\"{o}}m},\ and\
  \citenamefont {Eng}(2005)}]{Schneider2005}%
  \BibitemOpen
  \bibfield  {author} {\bibinfo {author} {\bibfnamefont {S.~C.}\ \bibnamefont
  {Schneider}}, \bibinfo {author} {\bibfnamefont {S.}~\bibnamefont
  {Grafstr{\"{o}}m}}, \ and\ \bibinfo {author} {\bibfnamefont {L.~M.}\
  \bibnamefont {Eng}},\ }\href@noop {} {\bibfield  {journal} {\bibinfo
  {journal} {Phys. Rev. B - Condens. Matter Mater. Phys.}\ }\textbf {\bibinfo
  {volume} {71}},\ \bibinfo {pages} {115418} (\bibinfo {year}
  {2005})}\BibitemShut {NoStop}%
\bibitem [{\citenamefont {Elings}\ \emph {et~al.}(1993)\citenamefont {Elings},
  \citenamefont {Q.}, \citenamefont {D.}, \citenamefont {K.}, \citenamefont
  {Zhong}, \citenamefont {Inniss}, \citenamefont {Kjoller},\ and\ \citenamefont
  {Elings}}]{Zhong1993}%
  \BibitemOpen
  \bibfield  {author} {\bibinfo {author} {\bibfnamefont {V.~B.}\ \bibnamefont
  {Elings}}, \bibinfo {author} {\bibfnamefont {Z.}~\bibnamefont {Q.}}, \bibinfo
  {author} {\bibfnamefont {I.}~\bibnamefont {D.}}, \bibinfo {author}
  {\bibfnamefont {K.}~\bibnamefont {K.}}, \bibinfo {author} {\bibfnamefont
  {Q.}~\bibnamefont {Zhong}}, \bibinfo {author} {\bibfnamefont
  {D.}~\bibnamefont {Inniss}}, \bibinfo {author} {\bibfnamefont
  {K.}~\bibnamefont {Kjoller}}, \ and\ \bibinfo {author} {\bibfnamefont
  {V.}~\bibnamefont {Elings}},\ }\href {\doibase 10.1016/0167-2584(93)90906-Y}
  {\bibfield  {journal} {\bibinfo  {journal} {Surf. Sci. Lett.}\ }\textbf
  {\bibinfo {volume} {290}},\ \bibinfo {pages} {688} (\bibinfo {year}
  {1993})}\BibitemShut {NoStop}%
\bibitem [{\citenamefont {Rakic}\ \emph {et~al.}(1998)\citenamefont {Rakic},
  \citenamefont {Djurisic}, \citenamefont {Elazar},\ and\ \citenamefont
  {Majewski}}]{Rakic1998}%
  \BibitemOpen
  \bibfield  {author} {\bibinfo {author} {\bibfnamefont {A.~D.}\ \bibnamefont
  {Rakic}}, \bibinfo {author} {\bibfnamefont {A.~B.}\ \bibnamefont {Djurisic}},
  \bibinfo {author} {\bibfnamefont {J.~M.}\ \bibnamefont {Elazar}}, \ and\
  \bibinfo {author} {\bibfnamefont {M.~L.}\ \bibnamefont {Majewski}},\ }\href
  {\doibase 10.1364/AO.37.005271} {\bibfield  {journal} {\bibinfo  {journal}
  {Appl. Opt.}\ }\textbf {\bibinfo {volume} {37}},\ \bibinfo {pages} {5271}
  (\bibinfo {year} {1998})}\BibitemShut {NoStop}%
\bibitem [{\citenamefont {G{\"{u}}thner}\ and\ \citenamefont
  {Dransfeld}(1992)}]{Guthner1992}%
  \BibitemOpen
  \bibfield  {author} {\bibinfo {author} {\bibfnamefont {P.}~\bibnamefont
  {G{\"{u}}thner}}\ and\ \bibinfo {author} {\bibfnamefont {K.}~\bibnamefont
  {Dransfeld}},\ }\href {\doibase 10.1063/1.107693} {\bibfield  {journal}
  {\bibinfo  {journal} {Appl. Phys. Lett.}\ }\textbf {\bibinfo {volume} {61}},\
  \bibinfo {pages} {1137} (\bibinfo {year} {1992})}\BibitemShut {NoStop}%
\bibitem [{\citenamefont {Eng}\ \emph {et~al.}(1999)\citenamefont {Eng},
  \citenamefont {G{\"{u}}ntherodt}, \citenamefont {Schneider}, \citenamefont
  {K{\"{o}}pke},\ and\ \citenamefont {{Mu{\~{n}}oz Salda{\~{n}}a}}}]{Eng1999}%
  \BibitemOpen
  \bibfield  {author} {\bibinfo {author} {\bibfnamefont {L.~M.}\ \bibnamefont
  {Eng}}, \bibinfo {author} {\bibfnamefont {H.-J.}\ \bibnamefont
  {G{\"{u}}ntherodt}}, \bibinfo {author} {\bibfnamefont {G.~A.}\ \bibnamefont
  {Schneider}}, \bibinfo {author} {\bibfnamefont {U.}~\bibnamefont
  {K{\"{o}}pke}}, \ and\ \bibinfo {author} {\bibfnamefont {J.}~\bibnamefont
  {{Mu{\~{n}}oz Salda{\~{n}}a}}},\ }\href {\doibase 10.1063/1.123266}
  {\bibfield  {journal} {\bibinfo  {journal} {Appl. Phys. Lett.}\ }\textbf
  {\bibinfo {volume} {74}},\ \bibinfo {pages} {233} (\bibinfo {year}
  {1999})}\BibitemShut {NoStop}%
\bibitem [{\citenamefont {Nonnenmacher}, \citenamefont {O'Boyle},\ and\
  \citenamefont {Wickramasinghe}(1991)}]{Nonnenmacher1991}%
  \BibitemOpen
  \bibfield  {author} {\bibinfo {author} {\bibfnamefont {M.}~\bibnamefont
  {Nonnenmacher}}, \bibinfo {author} {\bibfnamefont {M.~P.}\ \bibnamefont
  {O'Boyle}}, \ and\ \bibinfo {author} {\bibfnamefont {H.~K.}\ \bibnamefont
  {Wickramasinghe}},\ }\href {\doibase 10.1063/1.105227} {\bibfield  {journal}
  {\bibinfo  {journal} {Appl. Phys. Lett.}\ }\textbf {\bibinfo {volume} {58}},\
  \bibinfo {pages} {2921} (\bibinfo {year} {1991})}\BibitemShut {NoStop}%
\bibitem [{\citenamefont {Zerweck}\ \emph {et~al.}(2005)\citenamefont
  {Zerweck}, \citenamefont {Loppacher}, \citenamefont {Otto}, \citenamefont
  {Grafstr{\"{o}}m},\ and\ \citenamefont {Eng}}]{Zerweck2005}%
  \BibitemOpen
  \bibfield  {author} {\bibinfo {author} {\bibfnamefont {U.}~\bibnamefont
  {Zerweck}}, \bibinfo {author} {\bibfnamefont {C.}~\bibnamefont {Loppacher}},
  \bibinfo {author} {\bibfnamefont {T.}~\bibnamefont {Otto}}, \bibinfo {author}
  {\bibfnamefont {S.}~\bibnamefont {Grafstr{\"{o}}m}}, \ and\ \bibinfo {author}
  {\bibfnamefont {L.~M.}\ \bibnamefont {Eng}},\ }\href {\doibase
  10.1103/PhysRevB.71.125424} {\bibfield  {journal} {\bibinfo  {journal} {Phys.
  Rev. B - Condens. Matter Mater. Phys.}\ }\textbf {\bibinfo {volume} {71}},\
  \bibinfo {pages} {1} (\bibinfo {year} {2005})}\BibitemShut {NoStop}%
\bibitem [{\citenamefont {Shusterman}\ \emph {et~al.}(2007)\citenamefont
  {Shusterman}, \citenamefont {Raizman}, \citenamefont {Sher}, \citenamefont
  {Parltiel}, \citenamefont {Schwarzman}, \citenamefont {Lepkifker},\ and\
  \citenamefont {Rosenwaks}}]{Shusterman2007}%
  \BibitemOpen
  \bibfield  {author} {\bibinfo {author} {\bibfnamefont {S.}~\bibnamefont
  {Shusterman}}, \bibinfo {author} {\bibfnamefont {A.}~\bibnamefont {Raizman}},
  \bibinfo {author} {\bibfnamefont {A.}~\bibnamefont {Sher}}, \bibinfo {author}
  {\bibfnamefont {Y.}~\bibnamefont {Parltiel}}, \bibinfo {author}
  {\bibfnamefont {A.}~\bibnamefont {Schwarzman}}, \bibinfo {author}
  {\bibfnamefont {E.}~\bibnamefont {Lepkifker}}, \ and\ \bibinfo {author}
  {\bibfnamefont {Y.}~\bibnamefont {Rosenwaks}},\ }\href {\doibase
  10.1021/nl071031w} {\bibfield  {journal} {\bibinfo  {journal} {Nano Lett.}\
  }\textbf {\bibinfo {volume} {7}},\ \bibinfo {pages} {2089} (\bibinfo {year}
  {2007})}\BibitemShut {NoStop}%
\bibitem [{\citenamefont {Schumacher}\ \emph {et~al.}(2016)\citenamefont
  {Schumacher}, \citenamefont {Miyahara}, \citenamefont {Spielhofer},\ and\
  \citenamefont {Grutter}}]{Schumacher2016}%
  \BibitemOpen
  \bibfield  {author} {\bibinfo {author} {\bibfnamefont {Z.}~\bibnamefont
  {Schumacher}}, \bibinfo {author} {\bibfnamefont {Y.}~\bibnamefont
  {Miyahara}}, \bibinfo {author} {\bibfnamefont {A.}~\bibnamefont
  {Spielhofer}}, \ and\ \bibinfo {author} {\bibfnamefont {P.}~\bibnamefont
  {Grutter}},\ }\href {\doibase 10.1103/PhysRevApplied.5.044018} {\bibfield
  {journal} {\bibinfo  {journal} {Phys. Rev. Appl.}\ }\textbf {\bibinfo
  {volume} {5}},\ \bibinfo {pages} {1} (\bibinfo {year} {2016})}\BibitemShut
  {NoStop}%
\bibitem [{\citenamefont {Wurtz}, \citenamefont {Bachelot},\ and\ \citenamefont
  {Royer}(1998)}]{Wurtz1998}%
  \BibitemOpen
  \bibfield  {author} {\bibinfo {author} {\bibfnamefont {G.}~\bibnamefont
  {Wurtz}}, \bibinfo {author} {\bibfnamefont {R.}~\bibnamefont {Bachelot}}, \
  and\ \bibinfo {author} {\bibfnamefont {P.}~\bibnamefont {Royer}},\ }\href
  {\doibase 10.1063/1.1148834} {\bibfield  {journal} {\bibinfo  {journal} {Rev.
  Sci. Instrum.}\ }\textbf {\bibinfo {volume} {69}},\ \bibinfo {pages} {1735}
  (\bibinfo {year} {1998})}\BibitemShut {NoStop}%
\bibitem [{\citenamefont {Hillenbrand}, \citenamefont {Stark},\ and\
  \citenamefont {Guckenberger}(2000)}]{Hillenbrand2000}%
  \BibitemOpen
  \bibfield  {author} {\bibinfo {author} {\bibfnamefont {R.}~\bibnamefont
  {Hillenbrand}}, \bibinfo {author} {\bibfnamefont {M.}~\bibnamefont {Stark}},
  \ and\ \bibinfo {author} {\bibfnamefont {R.}~\bibnamefont {Guckenberger}},\
  }\href {\doibase 10.1063/1.126683} {\bibfield  {journal} {\bibinfo  {journal}
  {Appl. Phys. Lett.}\ }\textbf {\bibinfo {volume} {76}},\ \bibinfo {pages}
  {3478} (\bibinfo {year} {2000})}\BibitemShut {NoStop}%
\bibitem [{\citenamefont {Kehr}\ \emph {et~al.}(2017)\citenamefont {Kehr},
  \citenamefont {D{\"{o}}ring}, \citenamefont {Gensch}, \citenamefont {Helm},\
  and\ \citenamefont {Eng}}]{Kehr2017}%
  \BibitemOpen
  \bibfield  {author} {\bibinfo {author} {\bibfnamefont {S.~C.}\ \bibnamefont
  {Kehr}}, \bibinfo {author} {\bibfnamefont {J.}~\bibnamefont {D{\"{o}}ring}},
  \bibinfo {author} {\bibfnamefont {M.}~\bibnamefont {Gensch}}, \bibinfo
  {author} {\bibfnamefont {M.}~\bibnamefont {Helm}}, \ and\ \bibinfo {author}
  {\bibfnamefont {L.~M.}\ \bibnamefont {Eng}},\ }\href@noop {} {\bibfield
  {journal} {\bibinfo  {journal} {Synchrotron Radiat. News}\ }\textbf {\bibinfo
  {volume} {30}},\ \bibinfo {pages} {31} (\bibinfo {year} {2017})}\BibitemShut
  {NoStop}%
\bibitem [{\citenamefont {Chandler-Horowitz}\ and\ \citenamefont
  {Amirtharaj}(2005)}]{Chandler-Horowitz2005}%
  \BibitemOpen
  \bibfield  {author} {\bibinfo {author} {\bibfnamefont {D.}~\bibnamefont
  {Chandler-Horowitz}}\ and\ \bibinfo {author} {\bibfnamefont {P.~M.}\
  \bibnamefont {Amirtharaj}},\ }\href {\doibase 10.1063/1.1923612} {\bibfield
  {journal} {\bibinfo  {journal} {J. Appl. Phys.}\ }\textbf {\bibinfo {volume}
  {97}},\ \bibinfo {pages} {123526} (\bibinfo {year} {2005})}\BibitemShut
  {NoStop}%
\bibitem [{\citenamefont {Kischkat}\ \emph {et~al.}(2012)\citenamefont
  {Kischkat}, \citenamefont {Peters}, \citenamefont {Gruska}, \citenamefont
  {Semtsiv}, \citenamefont {Chashnikova}, \citenamefont {Klinkm{\"{u}}ller},
  \citenamefont {Fedosenko}, \citenamefont {Machulik}, \citenamefont
  {Aleksandrova}, \citenamefont {Monastyrskyi}, \citenamefont {Flores},\ and\
  \citenamefont {Masselink}}]{Kischkat2012}%
  \BibitemOpen
  \bibfield  {author} {\bibinfo {author} {\bibfnamefont {J.}~\bibnamefont
  {Kischkat}}, \bibinfo {author} {\bibfnamefont {S.}~\bibnamefont {Peters}},
  \bibinfo {author} {\bibfnamefont {B.}~\bibnamefont {Gruska}}, \bibinfo
  {author} {\bibfnamefont {M.}~\bibnamefont {Semtsiv}}, \bibinfo {author}
  {\bibfnamefont {M.}~\bibnamefont {Chashnikova}}, \bibinfo {author}
  {\bibfnamefont {M.}~\bibnamefont {Klinkm{\"{u}}ller}}, \bibinfo {author}
  {\bibfnamefont {O.}~\bibnamefont {Fedosenko}}, \bibinfo {author}
  {\bibfnamefont {S.}~\bibnamefont {Machulik}}, \bibinfo {author}
  {\bibfnamefont {A.}~\bibnamefont {Aleksandrova}}, \bibinfo {author}
  {\bibfnamefont {G.}~\bibnamefont {Monastyrskyi}}, \bibinfo {author}
  {\bibfnamefont {Y.}~\bibnamefont {Flores}}, \ and\ \bibinfo {author}
  {\bibfnamefont {W.~T.}\ \bibnamefont {Masselink}},\ }\href {\doibase
  10.1364/AO.51.006789} {\bibfield  {journal} {\bibinfo  {journal} {Appl.
  Opt.}\ }\textbf {\bibinfo {volume} {51}},\ \bibinfo {pages} {6789} (\bibinfo
  {year} {2012})}\BibitemShut {NoStop}%
\bibitem [{\citenamefont {Babar}\ and\ \citenamefont
  {Weaver}(2015)}]{Babar2015}%
  \BibitemOpen
  \bibfield  {author} {\bibinfo {author} {\bibfnamefont {S.}~\bibnamefont
  {Babar}}\ and\ \bibinfo {author} {\bibfnamefont {J.~H.}\ \bibnamefont
  {Weaver}},\ }\href {\doibase 10.1364/AO.54.000477} {\bibfield  {journal}
  {\bibinfo  {journal} {Appl. Opt.}\ }\textbf {\bibinfo {volume} {54}},\
  \bibinfo {pages} {477} (\bibinfo {year} {2015})}\BibitemShut {NoStop}%
\bibitem [{\citenamefont {Kehr}\ \emph {et~al.}(2008)\citenamefont {Kehr},
  \citenamefont {Cebula}, \citenamefont {Mieth}, \citenamefont
  {H{\"{a}}rtling}, \citenamefont {Seidel}, \citenamefont {Grafstr{\"{o}}m},
  \citenamefont {Eng}, \citenamefont {Winnerl}, \citenamefont {Stehr},\ and\
  \citenamefont {Helm}}]{Kehr2008}%
  \BibitemOpen
  \bibfield  {author} {\bibinfo {author} {\bibfnamefont {S.~C.}\ \bibnamefont
  {Kehr}}, \bibinfo {author} {\bibfnamefont {M.}~\bibnamefont {Cebula}},
  \bibinfo {author} {\bibfnamefont {O.}~\bibnamefont {Mieth}}, \bibinfo
  {author} {\bibfnamefont {T.}~\bibnamefont {H{\"{a}}rtling}}, \bibinfo
  {author} {\bibfnamefont {J.}~\bibnamefont {Seidel}}, \bibinfo {author}
  {\bibfnamefont {S.}~\bibnamefont {Grafstr{\"{o}}m}}, \bibinfo {author}
  {\bibfnamefont {L.~M.}\ \bibnamefont {Eng}}, \bibinfo {author} {\bibfnamefont
  {S.}~\bibnamefont {Winnerl}}, \bibinfo {author} {\bibfnamefont
  {D.}~\bibnamefont {Stehr}}, \ and\ \bibinfo {author} {\bibfnamefont
  {M.}~\bibnamefont {Helm}},\ }\href@noop {} {\bibfield  {journal} {\bibinfo
  {journal} {Phys. Rev. Lett.}\ }\textbf {\bibinfo {volume} {100}},\ \bibinfo
  {pages} {1} (\bibinfo {year} {2008})}\BibitemShut {NoStop}%
\bibitem [{\citenamefont {Reschke}\ \emph {et~al.}(2017)\citenamefont
  {Reschke}, \citenamefont {Mayr}, \citenamefont {Wang}, \citenamefont
  {Lunkenheimer}, \citenamefont {Li}, \citenamefont {Szaller}, \citenamefont
  {Bord{\'{a}}cs}, \citenamefont {K{\'{e}}zsm{\'{a}}rki}, \citenamefont
  {Tsurkan},\ and\ \citenamefont {Loidl}}]{Reschke2017}%
  \BibitemOpen
  \bibfield  {author} {\bibinfo {author} {\bibfnamefont {S.}~\bibnamefont
  {Reschke}}, \bibinfo {author} {\bibfnamefont {F.}~\bibnamefont {Mayr}},
  \bibinfo {author} {\bibfnamefont {Z.}~\bibnamefont {Wang}}, \bibinfo {author}
  {\bibfnamefont {P.}~\bibnamefont {Lunkenheimer}}, \bibinfo {author}
  {\bibfnamefont {W.}~\bibnamefont {Li}}, \bibinfo {author} {\bibfnamefont
  {D.}~\bibnamefont {Szaller}}, \bibinfo {author} {\bibfnamefont
  {S.}~\bibnamefont {Bord{\'{a}}cs}}, \bibinfo {author} {\bibfnamefont
  {I.}~\bibnamefont {K{\'{e}}zsm{\'{a}}rki}}, \bibinfo {author} {\bibfnamefont
  {V.}~\bibnamefont {Tsurkan}}, \ and\ \bibinfo {author} {\bibfnamefont
  {A.}~\bibnamefont {Loidl}},\ }\href {http://arxiv.org/abs/1704.08602}
  {\bibfield  {journal} {\bibinfo  {journal} {ArXiv e-prints}\ ,\ \bibinfo
  {pages} {1}} (\bibinfo {year} {2017})},\ \Eprint
  {http://arxiv.org/abs/1704.08602} {arXiv:1704.08602} \BibitemShut {NoStop}%
\bibitem [{\citenamefont {Hlinka}\ \emph {et~al.}(2016)\citenamefont {Hlinka},
  \citenamefont {Borodavka}, \citenamefont {Rafalovskyi}, \citenamefont
  {Docekalova}, \citenamefont {Pokorny}, \citenamefont {Gregora}, \citenamefont
  {Tsurkan}, \citenamefont {Nakamura}, \citenamefont {Mayr}, \citenamefont
  {Kuntscher}, \citenamefont {Loidl}, \citenamefont {Bord{\'{a}}cs},
  \citenamefont {Szaller}, \citenamefont {Lee}, \citenamefont {Lee},\ and\
  \citenamefont {K{\'{e}}zsm{\'{a}}rki}}]{Hlinka2016}%
  \BibitemOpen
  \bibfield  {author} {\bibinfo {author} {\bibfnamefont {J.}~\bibnamefont
  {Hlinka}}, \bibinfo {author} {\bibfnamefont {F.}~\bibnamefont {Borodavka}},
  \bibinfo {author} {\bibfnamefont {I.}~\bibnamefont {Rafalovskyi}}, \bibinfo
  {author} {\bibfnamefont {Z.}~\bibnamefont {Docekalova}}, \bibinfo {author}
  {\bibfnamefont {J.}~\bibnamefont {Pokorny}}, \bibinfo {author} {\bibfnamefont
  {I.}~\bibnamefont {Gregora}}, \bibinfo {author} {\bibfnamefont
  {V.}~\bibnamefont {Tsurkan}}, \bibinfo {author} {\bibfnamefont
  {H.}~\bibnamefont {Nakamura}}, \bibinfo {author} {\bibfnamefont
  {F.}~\bibnamefont {Mayr}}, \bibinfo {author} {\bibfnamefont {C.~A.}\
  \bibnamefont {Kuntscher}}, \bibinfo {author} {\bibfnamefont {A.}~\bibnamefont
  {Loidl}}, \bibinfo {author} {\bibfnamefont {S.}~\bibnamefont
  {Bord{\'{a}}cs}}, \bibinfo {author} {\bibfnamefont {D.}~\bibnamefont
  {Szaller}}, \bibinfo {author} {\bibfnamefont {H.~J.}\ \bibnamefont {Lee}},
  \bibinfo {author} {\bibfnamefont {J.~H.}\ \bibnamefont {Lee}}, \ and\
  \bibinfo {author} {\bibfnamefont {I.}~\bibnamefont {K{\'{e}}zsm{\'{a}}rki}},\
  }\href {\doibase 10.1103/PhysRevB.94.060104} {\bibfield  {journal} {\bibinfo
  {journal} {Phys. Rev. B}\ }\textbf {\bibinfo {volume} {94}},\ \bibinfo
  {pages} {1} (\bibinfo {year} {2016})}\BibitemShut {NoStop}%
\end{thebibliography}%

\end{document}